\documentclass[a4paper,11pt]{article}
\pdfoutput=1 
\usepackage{soul}
\usepackage{jcappub} 
\usepackage[T1]{fontenc} 
\usepackage[normalem]{ulem}
\usepackage{float}
\usepackage{lineno}
\usepackage[numbers,sort&compress]{natbib}
\usepackage{tikz}
\usetikzlibrary{arrows.meta}

\title{Catalog-level blinding on the bispectrum for DESI-like galaxy surveys}

\abstract{We evaluate the performance of the catalog-level blind analysis technique (\textit{blinding}) presented in Brieden et al. (2020) in the context of a  {fixed template} power spectrum and bispectrum analysis. This blinding scheme, which is tailored for galaxy redshift surveys similar to the Dark Energy Spectroscopic Instrument (DESI), has two components: the so-called ``AP blinding'' (concerning the dilation parameters $\alpha_\parallel,\alpha_\bot$) and ``RSD blinding''  (redshift space distortions, affecting the growth rate parameter $f$). Through extensive testing, including checks for the RSD part in cubic boxes, the impact of AP blinding on mocks with realistic survey sky coverage, and the implementation of a full AP+RSD blinding pipeline, our analysis demonstrates the effectiveness of the technique in preserving the integrity of cosmological parameter estimation when the analysis includes the bispectrum statistic. We emphasize the critical role of sophisticated---and difficult to accidentally unblind---blinding methods in precision cosmology.}

\author[1,2]{Sergi Novell-Masot,}
\author[1,2,3]{H\'ector Gil-Mar\'in,}
\author[1,4]{Licia Verde,}
\author[5]{{J.~Aguilar},}
\author[6]{{S.~Ahlen},}
\author[23]{{S.~Brieden},}
\author[7]{{D.~Brooks},}
\author[5]{{T.~Claybaugh},}
\author[9]{{A.~de la Macorra},}
\author[24,25]{{J.~E.~Forero-Romero},}
\author[2,11,12]{{E.~Gaztañaga},}
\author[5]{{S.~Gontcho A Gontcho},}
\author[26]{{G.~Gutierrez},}
\author[13,14,15]{{K.~Honscheid},}
\author[16]{{C.~Howlett},}
\author[17]{{R.~Kehoe},}
\author[5]{T.~Kisner,}
\author[5]{A.~Lambert,}
\author[5]{{M.~E.~Levi},}
\author[18,10]{{M.~Manera},}
\author[19]{{A.~Meisner},}
\author[2,10]{{R.~Miquel},}
\author[27,28]{{G.~Niz},}
\author[29]{{F.~Prada},}
\author[20]{{G.~Rossi},}
\author[21]{{E.~Sanchez},}
\author[8]{{M.~Schubnell},}
\author[22]{{H.~Seo},}
\author[19]{{D.~Sprayberry},}
\author[8]{{G.~Tarl\'{e}},}
\author[19]{{B.~A.~Weaver}}
\affiliation[1]{Institut de Ci\`encies del Cosmos (ICCUB), Universitat de Barcelona (UB), Mart\'i i Franqu\`es, 1, 08028 Barcelona, Spain}
\affiliation[2]{Institut d'Estudis Espacials de Catalunya (IEEC), 08034 Barcelona, Spain}
\affiliation[3]{Departament de F\'{\i}sica Qu\`{a}ntica i Astrof\'{\i}sica, Universitat de Barcelona, Mart\'{\i} i Franqu\`{e}s 1, E08028 Barcelona, Spain}
\affiliation[4]{Instituci\'{o} Catalana de Recerca i Estudis Avan\c{c}ats, Passeig de Llu\'{\i}s Companys, 23, 08010 Barcelona, Spain} 
\affiliation[5]{Lawrence Berkeley National Laboratory, 1 Cyclotron Road, Berkeley, CA 94720, USA}
\affiliation[6]{Physics Dept., Boston University, 590 Commonwealth Avenue, Boston, MA 02215, USA}
\affiliation[23]{Institute for Astronomy, University of Edinburgh, Royal Observatory, Blackford Hill, Edinburgh EH9 3HJ, UK}
\affiliation[7]{Department of Physics \& Astronomy, University College London, Gower Street, London, WC1E 6BT, UK}
\affiliation[8]{Department of Physics, University of Michigan, Ann Arbor, MI 48109, USA}
\affiliation[9]{{Instituto de F\'{\i}sica, Universidad Nacional Aut\'{o}noma de M\'{e}xico,  Cd. de M\'{e}xico  C.P. 04510,  M\'{e}xico}}
\affiliation[24]{Departamento de F\'isica, Universidad de los Andes, Cra. 1 No. 18A-10, Edificio Ip, CP 111711, Bogot\'a, Colombia}
\affiliation[25]{Observatorio Astron\'omico, Universidad de los Andes, Cra. 1 No. 18A-10, Edificio H, CP 111711 Bogot\'a, Colombia}
\affiliation[10]{Institut de F\'{i}sica d’Altes Energies (IFAE), The Barcelona Institute of Science and Technology, Campus UAB, 08193 Bellaterra Barcelona, Spain}
\affiliation[11]{Institute of Cosmology and Gravitation, University of Portsmouth, Dennis Sciama Building, Portsmouth, PO1 3FX, UK}
\affiliation[12]{Institute of Space Sciences, ICE-CSIC, Campus UAB, Carrer de Can Magrans s/n, 08913 Bellaterra, Barcelona, Spain}
\affiliation[26]{Fermi National Accelerator Laboratory, PO Box 500, Batavia, IL 60510, USA}
\affiliation[13]{Center for Cosmology and AstroParticle Physics, The Ohio State University, 191 West Woodruff Avenue, Columbus, OH 43210, USA}
\affiliation[14]{Department of Physics, The Ohio State University, 191 West Woodruff Avenue, Columbus, OH 43210, USA}
\affiliation[15]{The Ohio State University, Columbus, 43210 OH, USA}
\affiliation[16]{School of Mathematics and Physics, University of Queensland, 4072, Australia}
\affiliation[17]{Department of Physics, Southern Methodist University, 3215 Daniel Avenue, Dallas, TX 75275, USA}
\affiliation[18]{Departament de F\'{i}sica, Serra H\'{u}nter, Universitat Aut\`{o}noma de Barcelona, 08193 Bellaterra (Barcelona), Spain}
\affiliation[19]{NSF NOIRLab, 950 N. Cherry Ave., Tucson, AZ 85719, USA}
\affiliation[27]{Departamento de F\'{i}sica, Universidad de Guanajuato - DCI, C.P. 37150, Leon, Guanajuato, M\'{e}xico}
\affiliation[28]{Instituto Avanzado de Cosmolog\'{\i}a A.~C., San Marcos 11 - Atenas 202. Magdalena Contreras, 10720. Ciudad de M\'{e}xico, M\'{e}xico}
\affiliation[29]{Instituto de Astrof\'{i}sica de Andaluc\'{i}a (CSIC), Glorieta de la Astronom\'{i}a, s/n, E-18008 Granada, Spain}
\affiliation[20]{Department of Physics and Astronomy, Sejong University, Seoul, 143-747, Korea}
\affiliation[21]{CIEMAT, Avenida Complutense 40, E-28040 Madrid, Spain}
\affiliation[22]{Department of Physics \& Astronomy, Ohio University, Athens, OH 45701, USA}
\emailAdd{sergi.novell@icc.ub.edu, hectorgil@icc.ub.edu, liciaverde@icc.ub.edu}

\begin{document}
\maketitle

\section{Introduction}

In recent decades, cosmology has undergone remarkable progress, entering the era of precision cosmology, with the $\Lambda$CDM model as a cornerstone, bridging theory and observations throughout all observable epochs of the Universe. However, there are still several open questions at the heart of our understanding of cosmology. We still do not know the fundamental physics of dark matter and dark energy, which amount to roughly 95\% of the Universe, and some tensions arise from the $\Lambda$CDM model when applied to both early and late-time physics. Current and forthcoming experiments---both early-time, such as the Southern Pole Telescope (SPT) \cite{benson2014spt}, the Atacama Cosmology Telescope (ACT) \cite{aiola2020atacama}, the Simons Observatory \cite{galitzki2018simons}, as well as late-time, such as the Dark Energy Spectroscopic Instrument (DESI) \cite{Snowmass2013.Levi,DESI2016a.Science,DESI2016b.Instr,DESI2022.KP1.Instr,FocalPlane.Silber.2023,Corrector.Miller.2023,Spectro.Pipeline.Guy.2023,SurveyOps.Schlafly.2023,LRG.TS.Zhou.2023,DESI2023a.KP1.SV,DESI2023b.KP1.EDR,2024arXiv240403000D,2024arXiv240403001D,2024arXiv240403002D}, Euclid \cite{laureijs2011euclid}, Vera Rubin \cite{ivezic2019lsst}---will play a pivotal role in these unresolved issues, yielding complementary cosmological information with an unprecedented level of precision. 

In order to ensure the reliability of results and mitigate confirmation bias, blind analyses, whereby the true result is hidden to the experimenter until the full analysis is done and ``frozen'', have been increasingly adopted in cosmology. Blind analyses, already standard practice in fields like experimental particle physics \cite{klein2005blind}, have been extended to cosmology in recent years \cite{conley2006measurement,maccoun2015blind,kuijken2015gravitational,zhang2017blinded,abbott2018dark,asgari2020kids+}. However, there is no one-size-fits-all standard for making an analysis blind; hence there is a need to develop and use a suitable method for each context. Hereafter, we refer to the action of making an analysis blind as simply ``blinding'', while we denote as ``unblinding'' the procedure of recovering the original, non-blind results. 

One can organize the different approaches to blinding relative to the stage of the analysis in which they are implemented. Generally, in large-scale structure surveys, the earlier it is implemented the more difficult it is to accidentally unblind. For example, while adding a blinding scheme just at the final analysis stages, e.g. shifting the recovered cosmological parameters by an unknown amount, is certainly convenient, it is nevertheless relatively easy to accidentally unblind. An intermediate approach involves adding a random shift in the summary statistics or their covariance \cite{muir2020blinding,sellentin2020blinding,fontribera2024inprep}, e.g. shifting the wave-vectors in the power spectrum. The strongest methods are generally the ones that are implemented already at the phase of catalog production, as done for example in \cite{asgari2020kids+}. 

In the framework of large-scale structure analyses, focusing on the power spectrum baryon acoustic oscillations (BAO) feature and redshift-space distortions, the method developed by \cite{brieden2020blind}
operates at catalog-level, which is currently part of the official DESI pipeline \cite{andrade2024validating}. While this method has been extensively validated in the context of a power spectrum template-based approach \cite{andrade2024validating,2024arXiv240403000D}, it still has not been tested on higher-order statistics such as the bispectrum, which is what we set out to do. The purpose of this work is to validate the blinding strategy of \cite{brieden2020blind} for an analysis featuring the combination of the power spectrum and bispectrum multipoles in DESI-like periodic boxes and cutsky mocks.

In Section~\ref{sec: theory_blinding} we review the main aspects of this blinding procedure. In Section~\ref{sec: 3} we describe the simulations and the main ingredients of our analysis. In Section~\ref{sec: results} we present our results and we conclude in Section~\ref{sec: conclusions}.

\section{Blinding the data at the catalog level}
\label{sec: theory_blinding}

The full blinding procedure consists of two separate steps, which we refer to as respectively AP and RSD parts of the blinding, following the nomenclature from \cite{brieden2020blind}. It was shown in \cite{brieden2020blind} that this combination results in a modular effect in the cosmological parameters (with the AP and RSD parts affecting different sets of cosmological parameters) while being very difficult to accidentally unblind. In particular, the RSD blinding modifies the linear growth rate parameter $f$, while the AP part of the blinding results in shifts in the dilation parameters $\{\alpha_\parallel,\alpha_\bot\}$, which are respectively the parallel and perpendicular to the line of sight dilation scales with respect to the fiducial cosmology. At any redshift $z$, given the Hubble parameter $H(z)$, the sound horizon scale at baryon drag redshift $r_\textrm{s}(z_\textrm{d})$ and the angular distance parameter $D_\textrm{A}(z)$, the dilation parameters are defined as  \cite{Alcock:1979mp}
\begin{equation}
\alpha_\parallel(z)=\frac{H^\textrm{fid}(z)r_\textrm{s}^\textrm{fid}(z_\textrm{d})}{H(z)r_\textrm{s}(z_\textrm{d})};\quad \alpha_\bot(z)=\frac{D_\textrm{A}(z)r_\textrm{s}^\textrm{fid}(z_\textrm{d})}{D_\textrm{A}^\textrm{fid}(z)r_\textrm{s}(z_\textrm{d})},
\end{equation}
with the ``fid'' superscript referring to the parameter corresponding to the fiducial cosmology.

In this section, we provide an updated summary of the procedure of \cite{brieden2020blind}. In a nutshell, the AP component of the blinding acts only along the line of sight by hiding (blinding) the reference cosmology used to convert the redshifts ($z$) into comoving distances. It does so by converting the original redshifts (indicated by $z$) of the catalog to distances using a hidden  (blind or shifted in \cite{brieden2020blind} nomenclature) cosmology and then transforming back the distances into redshifts using a reference (known) cosmology, thus producing an effectively blind redshift catalog. Blind redshifts are indicated by $z'$. It is easy to see that this transformation only affects the dilation 
parameters $\alpha_\parallel,\,\alpha_\bot$.

Let us refer to the vector of cosmological parameters that completely specify a given cosmological model by ${\bf \Omega}$. For example, in a standard $\Lambda$CDM model, ${\bf \Omega}$ includes parameters such as  the matter density parameter $\Omega_m$, the baryon density parameter $\Omega_b$, the primordial power spectrum spectral slope $n_s$, the Hubble constant $H_0$, the rms dark matter fluctuations filtered on $8$ Mpc/$h$ scales $\sigma_8$.
The blinding scheme that the DESI collaboration  uses shifts a given  reference cosmology ${\bf \Omega}^{\rm ref}$ by a (hidden) shift $\Delta {\bf \Omega}$ such that a blind cosmology is obtained:
\begin{equation}
{\bf \Omega}'={\bf \Omega}^{\rm  ref}+\Delta{\bf \Omega}\equiv {\bf \Omega}^{\rm ref}+\left(\Delta f,\Delta w_0,\Delta w_a\right)
\label{eq:shiftsparams}
\end{equation} 
where the linear growth rate is taken to be $f=\Omega_m^\gamma$,\footnote{Here, the exponent $\gamma$ is chosen following the General Relativity prescription, i.e. $\gamma\approx 6/11$.\cite{linder2005cosmic,linder2007parameterized}} and $w_0,w_a$  describe dynamical dark energy with dark energy equation of state parameter being a function of the scale factor: $w(a) = w_0+w_a(1-a)$. 
When the analysis on the blind data is performed, a fiducial cosmology ${\bf \Omega}^{\rm fid}$ is used (hence transforming equation \ref{eq:shiftsparams} such that ${\bf \Omega}^{\rm ref}\rightarrow {\bf \Omega}^{\rm fid}$). 

 {We should note here that the reference cosmology used to produce the blinding catalog is not necessarily the same as the fiducial cosmology used for generating the fixed template employed for fitting the measurements and performing cosmological inference on the catalog. In summary, in this paper we refer to the cosmology assumed when generating the unblinded catalogue in cartesian coordinates as the ``reference'', and the associated blind parameters with respect to the reference are marked with a prime ('). Additionally, we refer to the cosmology assumed later in performing the cosmological analysis and measurement of the blind catalog as ``fiducial'', whereas the associated recovered parameters are notated with the superscript ``blind''. The prime (') and the ``blind'' quantities generally  do not coincide.  This is sketched in the following diagram:\vspace{.5cm}}

\begin{tikzpicture}[auto]

    \node (unblinded) at (0,0) {Unblinded catalog};
    \node (blinded) at (5.5,0) {Blinded catalog};
    \node (omega_ref) at (2.7,-1.) {$\Omega^{\text{ref}}, \Omega'$};
    \node (omega_fid) at (7.6,-1.) {$\Omega^{\text{fid}}$};
    \node (omega_blind) at (10,0) {$\Omega^{\text{blind}}$};
    
    \draw[-{Stealth}] (unblinded) -- (blinded);
    \draw[-{Stealth}] (blinded) -- (omega_blind);
    \draw[-{Stealth}] (omega_ref) -- (blinded);
    \draw[-{Stealth}] (omega_fid) -- (omega_blind);
    
\end{tikzpicture}

Since in this specific application we do not use real data or an unknown true cosmology, the prime quantities are the blinded ones, while the non prime quantities are the reference (and true) ones. 

The resulting shifts from Equation \ref{eq:shiftsparams} in the  dilation 
parameters are:
\begin{equation}
\label{eq: alphas}
    \frac{\alpha_\parallel'(z')}{\alpha^\textrm{ref}_\parallel(z)}=\frac{H'(z)}{H^\textrm{ref}(z)};\quad\quad \frac{\alpha_\bot'(z')}{\alpha^\textrm{ref}_\bot(z)}=\frac{D_M^\textrm{ref}(z)}{D'_M(z)},
\end{equation}
  with $H$ and $D_M$ being respectively the Hubble expansion parameter and the angular comoving distance at redshift $z$, 
\begin{align}
    H(z)&=H_0\sqrt{\Omega_m(1+z)^3+(1-\Omega_m)(1+z)^{3(1+w_0+w_a(1-a(z)))}}\\ 
    D_M(z)&=\int_0^zdz'\frac{c}{H(z')}.
\end{align}
The dilation parameters expected to be measured from the blinded catalog $\alpha^{\rm blind}_{\parallel,\perp}$ are related to the pre-blinding ones  $\alpha_{\parallel,\perp}$ by the same ratio as the $\alpha'$'s to the  $\alpha^{\rm ref}$s.  {In other words, when the blinded catalog is analyzed, by using a fiducial cosmology, Equation \ref{eq: alphas}  holds, with the modification that the prime quantities are notated as ``blind'' and the superscript ``ref'' is changed by ``fid''.} Of course, in tests applied to simulations the reference quantities coincide with the fiducial and true quantities, and the prime coincides with the blind, i.e.   ${\bf \Omega}^{\rm ref}={\bf \Omega}^{\rm fid}={\bf \Omega}^{\rm true}$ and ${\bf \Omega}'={\bf \Omega}^{\rm blind}$.

The RSD part of the blinding operates in a different way. 
Built on the RSD part of the reconstruction algorithm \cite{eisenstein2007robustness,padmanabhan2009reconstructing,burden2014efficient}, this step of the blinding procedure smooths the density field with a Gaussian filter 
and then uses the reconstructed field to shift the redshift space positions $\textbf{r}$ according to,

\begin{equation}
    \mathbf{r}'=\mathbf{r}-f^\textrm{ref}(\Psi\cdot\hat{\mathbf{r}})\hat{\mathbf{r}}+ f'\textrm(\Psi\cdot \hat{\mathbf{r}})\hat{\mathbf{r}}.
\end{equation}
The quantity $\Psi\cdot \hat{\mathbf{r}}$, where $\Psi$ is the displacement field\footnote{ {The displacement field is obtained with \href{https://github.com/desihub/LSS/blob/main/scripts/recon.py}{the reconstruction algorithm}, by assuming reference values for the growth factor $f$ and galaxy bias $b$. It has been shown that a reasonable choice of the reference cosmology in the reconstruction algorithm will not introduce neither significant systematic errors \cite{carter2020impact,bernal2020robustness,sherwin2019impact} nor error mis-estimation \cite{brieden2020blind}.}}.
Following \cite{brieden2020blind} this is  inferred from the reconstructed real-space positions of the reconstructed field.  {In  fact, the difference between original and reconstructed comoving distances is equal to the difference in growth factors times the displacement along the line of sight:}
\begin{equation}
    d^\textrm{ref}-d^\textrm{rec}=(f^\textrm{ref}-f^\textrm{rec})(\Psi\cdot\hat{\mathbf{r}})\hat{\mathbf{r}}=f^\textrm{ref}(\Psi\cdot\hat{\mathbf{r}})\hat{\mathbf{r}},
\end{equation}
 {since $f^\textrm{rec}=0$. Applying the same relationship to the difference in distances between the blinded and reference distances gives us}
\begin{equation}
    d'-d^\textrm{ref}=(f'-f^\textrm{ref})(\Psi\cdot\hat{\mathbf{r}})\hat{\mathbf{r}}=(f'-f^\textrm{ref})\frac{d^\textrm{ref}-d^\textrm{rec}}{f^\textrm{ref}},\\
\end{equation}
with $d^\textrm{rec},d^\textrm{ref}$ being respectively the reconstructed and reference comoving distances, and $f',f^\textrm{ref}$ the shifted and reference values of the growth factor $f$. Rearranging the terms, we obtain the blinded comoving distances as 
\begin{equation}
\label{eq: RSD_distance}
    d'=d^\textrm{rec}+\frac{f'}{f^\textrm{ref}}(d^\textrm{ref}-d^\textrm{rec}).
\end{equation}
 It should now be clear that in this way, the redshifts of the tracers are modified such that the catalog has effectively different RSD signal, of amplitude given by $f'=f^{\rm true}+\Delta f$.

Both parts of the blinding procedure have been tested to be robust with a template-based power spectrum analysis \cite{brieden2020blind,andrade2024validating} for an extensive set of possible blinding shifts, and we set to quantify their performance (both separately and in combination) for analyses involving the bispectrum. 

\section{Test design and analysis set-up}
\label{sec: 3}
\subsection{Test design and current limitations}
In this paper we perform the following sequence of tests for a joint power spectrum and bispectrum analysis which employs cubic (periodic) boxes and more realistic mock surveys which include partial sky coverage  (cutsky mocks):\footnote{While cubic mocks simulate the cosmic structure within a periodic cubic volume, cutsky mocks apply observational features to these simulations to mimic real sky survey data, accounting for survey geometry and selection effects.}
\begin{enumerate}
    \item Performance of the RSD part with cubic boxes for the combination of the power spectrum and bispectrum redshift space multipoles.
    \item Performance of the AP part with cutsky mocks for the combination of redshift space power spectrum multipoles and bispectrum monopole.
    \item Performance of the full AP+RSD blinding with cutsky mocks for the combination of redshift space power spectrum multipoles and bispectrum monopole.
\end{enumerate}

The reason why we first test the RSD part of the blinding algorithm in cubic boxes is that the RSD part of the algorithm could be more sensitive than the AP part to the specific analysis settings.
Implementing the AP part of the blinding (which was designed with cutsky data in mind) in a cubic box is challenging both conceptually and practically. First of all, AP blinding cannot be done correctly in plane-parallel approximation (unless in the limit of large distances and small survey areas, which is not the case here). Secondly,  the advantage of using cubic boxes is their periodicity, but periodicity is lost when applying the AP blinding.  For more details see  Section \ref{subsec: sims}.  
Given this, we separately test the AP part of the blinding in step 2.

To perform step 1, we modify the blinding code to be used by DESI in order to apply the RSD part of the blinding to cubic boxes.\footnote{\url{https://github.com/desihub/LSS/}} 
For this test, we use the global plane-parallel approximation by obtaining the redshifts of the galaxies from the third direction of the box, $\hat{x}_3$. Then, the first two coordinates  $(x_1,x_2)$ are left equal, while the third coordinate $x_3$ is transformed according to Equation \ref{eq: RSD_distance}\footnote{If the transformed value for $x_3$, which we can denote as $\Tilde{x}_3$, falls outside of the box, it is reintroduced assuming periodic boundary conditions as $\Tilde{x}_3$ modulus $L_\textrm{box}$}.

In this work,  for steps 2 and 3  we limit ourselves to adding only the bispectrum monopole to the power spectrum multipoles. The currently available treatment of the bispectrum window function \cite{Gil-Marin:2014biasgravity,Gil-Marin:2016sdss}
is only applicable to the bispectrum monopole $B_0$. We recognize that this bispectrum window function treatment is known to suffer from some degree of inaccuracy, especially in the squeezed configurations \cite{pardede2022bispectrum};  as we will show, these do not interfere with the performance of the blinding  (and unblinding) procedure.

We leave to future work to update steps 2 and 3 to include also the bispectrum quadrupoles, subject to the availability of an improved window function implementation suitable for the bispectrum multipoles.

\subsection{Simulations and blinding parameters}
\label{subsec: sims}
We utilize 25 \textsc{AbacusSummit} LRG simulations at redshift $z=0.8$, which we may refer to as simply as the \textsc{AbacusSummit} LRG mocks \cite{maksimova2021abacussummit}.  LRG stands for Luminous Red Galaxies, which constitutes the highest signal-to-noise  DESI sample at $z\lesssim 0.8$ \cite{LRG.TS.Zhou.2023}. For the first test in this work (RSD blinding, as presented in Section \ref{sec: theory_blinding}) we use the 25 available periodic boxes, while for the second and third tests, we use the corresponding 25 cutsky mocks. The sky area and completeness of the window that we use in cutsky mocks coincide with the DESI Year 1 footprint, and we measure the power spectra and bispectra using random catalogs (randoms) with 20 times the density of the data.

The cosmology of all the mocks as well as our adopted reference and fiducial cosmology are the same and compatible with the \textsc{Planck} best fit $\Lambda$CDM model \cite{collaboration2018planck}: 
$\Omega_m=0.3138,\Omega_b=0.0493,$ 
$n_\textrm{s}=0.9649,\sigma_8=0.8114,h=0.6736,w=-1.$ The cubic boxes have a size of $L_\textrm{box}=2000\,\textrm{Mpc}\,h^{-1}$, resulting in a total physical volume for the 25 boxes of $V_{25}=200\,(\textrm{Gpc}/h)^3$. The effective volume of a single cubic box---computed as in \cite{tegmark1997measuring}---is $\sim6.4\,(\textrm{Gpc}\,h^{-1})^3$, for a total effective volume of $V^{\rm eff}_{25}\sim 160\, (\textrm{Gpc}\,h^{-1})^3$. The cutsky mocks are generated by the Generate Survey Mocks code\footnote{\url{https://github.com/Andrei-EPFL/generate_survey_mocks}} \cite{variuinprep, white2014mock} 
and matched the DESI Year 1 survey footprint with the mkfast-Y1 code\footnote{\url{https://github.com/desihub/LSS/blob/main/scripts/mock_tools/mkfast_Y1.py}}. 
The effective volume of each cutsky mock is estimated to be $\sim 1.6 (\textrm{Gpc}\,h^{-1})^3$ resulting in the 25 cutsky mocks having a cumulative effective volume\footnote{The effective volume of each cutsky mock is calculated by multiplying the effective volume of its correspondent cubic mock ($\sim6.4\,(\textrm{Gpc}\,h^{-1})^3$ \cite{tegmark1997measuring}) by the ratio between the number of particles in the cutsky mock and the cubic box.} of $V^{\rm eff}_{\rm 25cutsky}\sim40\,(\textrm{Gpc}\,h^{-1})^3$, of the order of the expected DESI final volume (5 years of survey) for LRGs \cite{LRG.TS.Zhou.2023}.

From the original \textsc{AbacusSummit} suite, we produce a total of 75 blinded mocks, obtained by applying to each original \textsc{AbacusSummit} LRG simulation the corresponding blinding as per Section \ref{sec: theory_blinding} (25 blinded mocks for each of the 3 tests). 
We blind the cubic boxes by a different amount than the cutsky mocks.
The blinding is such that the expected shift in the growth rate $f$ is $\Delta f=f'-f^\textrm{ref}=-0.068$ for the cubic boxes (where the value of $f^\textrm{ref}$ is set to 0.8), while for the cutsky mocks we change the sign of the expected shift so that $\Delta f=f'-f^\textrm{ref}=0.060$.  The AP part of the blinding in the cutsky mocks has expected shifts of $\Delta \alpha_\parallel=\Delta \alpha_\bot=-0.013$.\footnote{Note that in general the $\alpha_\parallel$ and $\alpha_\bot$ parameters are not necessarily shifted equally. The fact that in our case the shift is the same is just a coincidence.}  {We only consider one blinding shift for the $\alpha_\parallel,\alpha_\bot$ parameters given the fact that the AP part of the blinding is based on purely geometrical arguments. In this case, the effect of the AP blinding can be  described analytically and  affects all correlations coherently, as it has been extensively tested in \cite{brieden2020blind}. }

The values of the shifts were chosen in order to be close to the limit of the DESI blinding pipeline, which enforces $|\Delta f|\leq 0.08$ $|\Delta \alpha_\parallel|\leq 0.03$, $|\Delta \alpha_\bot|\leq 0.03$ \cite{andrade2024validating}. Compared to the expected standard deviation of the parameters recovered from the total volume of the 25 mocks ($V_{25}$), these offsets 
represent a blinding shift in $f$ for the cubic boxes of $\sim3\sigma$,  and for the cutsky mocks of  $\sim1\textrm{--}2\sigma$ for all the parameters  $f,\alpha_\parallel,\alpha_\bot$. We choose not to go beyond $3\sigma$ of the expected errorbars in order to not produce a degradation in the recovered constraints \cite{brieden2020blind}.

It should now be clearer why we do not implement the AP blinding on cubic boxes. The size of the cubic boxes ($L_\textrm{box}=2000\,\textrm{Mpc}\,h^{-1}$) is of comparable magnitude to the comoving distance from $z=0$ to $z=0.8$ where the tracers are located. Hence, to introduce a significant  AP shift at $z=0.8$ would require arbitrarily setting the observer at a relatively short distance from the box. This would break periodicity and the plane parallel approximation for RSD would not apply. This limitation does not apply to the cutsky mocks.

For estimating the covariance matrices, we measure the power spectra and bispectra from 1000 EZmocks, which generate the density field using the effective Zel'dovich approximation \cite{chuang2015ezmocks}. Both the underlying cosmology and the tracers' clustering properties are designed to match the \textsc{AbacusSummit} LRG mocks. Throughout, we measure the power spectrum and bispectrum of both cubic and cutsky mocks using the publicly available \textsc{Rustico}  code\footnote{\url{https://github.com/hectorgil/Rustico}} \cite{HGMeboss}.

\subsection{Modeling}
Our theoretical modelling is the one adopted in our previous works \cite{novell2023geofpt,novell2023approximations}. The interested reader can refer to such works {, or to Appendix \ref{app: theory},} to find an in-depth description  {of the theoretical modelling}. The modelling of the power spectrum follows the renormalized perturbation theory (RPT) prescription \cite{Crocce_2006,gil-marin_power_2015}. For the bispectrum 
we use the phenomenological GEO-FPT model \cite{novell2023geofpt}, as provided in  the publicly available GEO-FPT code\footnote{\url{https://github.com/serginovell/Geo-FPT}}. In particular for this application, the phenomenological parameters that model the bispectrum shape and scale dependence, $\{f_1,...,f_5\}$,  are obtained at $z=0.8$  by interpolation from the tabulated values provided by GEO-FPT.

In all cases we use the power spectrum monopole and quadrupole data-vector, $P_{02}=\{P_0,P_2\}$, while we consider different parts of the bispectrum multipole expansion according to the analysis settings: for cubic mocks (Section \ref{sec: cubic}) we use the bispectrum monopole together with the first two quadrupoles,\footnote{We found in \cite{novell2023geofpt} that it is redundant to use all three quadrupoles $B_{200},B_{020},B_{002}$. We follow the bispectrum multipole expansion convention by \cite{Scoccimarro:1999ed}.} and refer to our full bispectrum data-vector as $B_{02}=\{B_0,B_{200},B_{020}\}$; for cutsky mocks (Section \ref{sec: cutsky}) we only consider the bispectrum monopole $B_0$.

As in our previous works \cite{novell2023geofpt,novell2023approximations}, we perform the cosmological parameter inference via a Markov chain Monte Carlo sampling (MCMC), specifying broad uniform priors in all parameters.  {The MCMC is performed using the \textsc{Brass}\footnote{\url{https://github.com/hectorgil/Brass}} \cite{HGMeboss} code, which implements a Metropolis-Hasting algorithm and tests the convergence via the Gelman-Rubin convergence criteria, with $R<1.01$. We specify broad, uniform priors---which are effectively uniform improper priors since the MCMC never samples the prior boundary---in all parameters except from $A_P,A_B$, whose prior is $\mathcal{N}(1,0.3)$.} We use the full shape of the power spectrum and bispectrum,  {using a fixed template approach,} under the assumption of local Lagrangian bias \cite{baldauf2012evidence,saito2014understanding,Brieden_ptchallenge}. The parameters of interest in this work are $\{f,\sigma_8,\alpha_\parallel,\alpha_\bot\}$, and we marginalize over the nuisance parameters $\{b_1,b_2,A_\textrm{P},A_\textrm{B},\sigma_\textrm{P},\sigma_\textrm{B}\}$, respectively: the linear and quadratic bias parameters, the shot noise amplitude correction for the power spectrum and bispectrum, and the power spectrum and bispectrum peculiar velocities rms.

 {The $k$-range that we consider for the power spectrum is of $0.02<k [h\,\textrm{Mpc}^{-1}]<0.15$, with a bin size of $\Delta k=0.01\,h\,\textrm{Mpc}^{-1}\approx 3.2k_\textrm{f}$,}\footnote{ {The fundamental wave-vector, $k_\textrm{f}$, is defined as $k_\textrm{f}=2\pi/L_\textrm{box}$.}}. The main difference with the analysis presented in previous works is that we limit the bispectrum $k$-range to $0.02\,h{\rm Mpc}^{-1}<k<0.11\,h{\rm Mpc}^{-1}$ (the bispectrum model was calibrated for $k_\textrm{max}=0.12\,h{\rm Mpc}^{-1}$). The reason for this choice is that the size of the data-vector is limited by the available number of simulations (1000 EZmocks). In order to estimate the full covariance matrix (as recommended in \cite{novell2023approximations}) the number of simulations should be much bigger than  
the size of the data-vector \cite{Hartlap:2006kj}. 
By adopting this $k_\textrm{max}$ and the relatively large binning size of $\Delta k^B=0.01\,h\,\textrm{Mpc}^{-1}$ the full $P_{02}$+$B_{02}$ data-vector size is of 342 elements, while the $P_{02}$+$B_0$ data-vector used in the cutsky mocks has 152 elements. We further account for  the errors due to the finite number of simulations in the covariance matrix estimation by employing the Sellentin-Heavens likelihood function \cite{Sellentin:2015waz}.

Finally, for the cases where we use cutsky mocks, we model the survey window for the power spectrum and bispectrum exactly as in \cite{Gil-Marin:2014biasgravity,Gil-Marin:2016sdss}. In short, the galaxy power spectrum model $P_\textrm{gal}$ is convolved with the window function $W_2$ (obtained by performing pair counts in the random catalog, as defined in \cite{Gil-Marin:2014biasgravity,Gil-Marin:2016sdss}) to obtain the windowed power spectra $P^{W}$ as
\begin{equation}
\label{eq: win_pk}
    P^W(\textbf{k})=\int\frac{d^3\textbf{k}'}{(2\pi)^3}P_\textrm{gal}(\textbf{k}')|W_2(\textbf{k}-\textbf{k}')|^2+P_\textrm{noise}(\textbf{k}).
\end{equation}
where $P_{\rm noise}$ denotes the shot noise contribution to the power spectrum.

Due to the computational challenge of operating with the analogous expression of Equation \ref{eq: win_pk} in the case of the bispectrum, in this paper we use the approximation of \cite{Gil-Marin:2014biasgravity,Gil-Marin:2016sdss}, which  ignores the effect of the window on the bispectum kernel. 
In fact, we can factorize the model of the galaxy bispectrum as $B_\textrm{gal}(\textbf{k}_1,\textbf{k}_2,\textbf{k}_3)=P_{NL}(\textbf{k}_1)P_{NL}(\textbf{k}_2)\mathcal{Z}(\textbf{k}_1,\textbf{k}_2,\textbf{k}_3)+\textrm{cyc.}$, where $P_{NL}$ is the non-linear matter power spectrum, and the function $\mathcal{Z}(\textbf{k}_1,\textbf{k}_2,\textbf{k}_3)$ encompasses the perturbation theory kernels, the geometrical correction and the bias expansion. In our adopted approximation, the $\mathcal{Z}$ function is unaffected by the window, resulting in a windowed bispectrum monopole theory vector $B^W$ that reads
\begin{equation}
\label{eq: win_B}
    B^W_0(k_1,k_2,k_3)=\int_{-1}^{1}d\mu_1\int_0^{2\pi}d\phi \mathcal{Z}(\textbf{k}_1,\textbf{k}_2,\textbf{k}_3)P_{NL}^W(\textbf{k}_1)P_{NL}^W(\textbf{k}_2)+\textrm{cyc}.
\end{equation}
In the above expression, $\mu_1$ is the cosine of the angle of the vector $\textbf{k}_1$ with respect to the line of sight, and $\phi$ is such that $\mu_2=\mu_1\cos\theta_{12}-\sqrt{(1-\mu_1^2)(1-\cos\theta_{12}^2)}\cos\phi$, with $\theta_{12}$ being the angle between $k_1$ and $k_2$.
Since this approximation  cannot  be easily generalized to the bispectrum quadrupoles, and to date there is no suitable window modelling for the bispectrum quadrupoles in this particular multipole expansion, 
in this work for cutsky tests we only use the bispectrum monopole.\footnote{As a consequence, in the cutsky cases, where the bispectrum quadrupoles are not used, we do not consider variations from Poisson shot-noise in the bispectrum, as we do with the full joint power spectrum and bispectrum multipoles analysis via the $A_P$ and $A_B$ parameters respectively. We find that letting $A_B$ vary when only using the bispectrum monopole $B_0$ opens up an unnecessarily large and unconstrained degeneracy direction where cosmological parameters take unphysical values, e.g. $f>1$.}  
For the purpose of testing the blinding, we find the current approximation of Equation \ref{eq: win_B} to be sufficient.
 To show that the adopted model for the data vector does not display obvious signature of systematics,  Appendix \ref{sec: app} shows the best fitting theoretical model for the data vector  from the blind and not blind catalogs, along with relevant $\chi^2$ values.

\section{Results} 
\label{sec: results}

In order to test the validity of the RSD part of the blinding scheme for the addition of the bispectrum in the data-vector, we perform similar tests as in \cite{brieden2020blind}, while using the set-up and analysis previously done in \cite{novell2023geofpt} and adapted as mentioned in Section \ref{sec: 3}.

\subsection{Cubic mocks: RSD blinding}
\label{sec: cubic}

We start by running the cosmological pipeline using only the power spectrum monopole and quadrupole $P_{02}$ on the 25 original and RSD-blinded \textsc{AbacusSummit} mocks. Additionally, we repeat the procedure for the mean of the 25 simulations, taking advantage of the fact that we have blinded all realizations with the same shift $\Delta f$. 
\begin{figure}[ht!]
\centering 
\includegraphics[width = \textwidth]{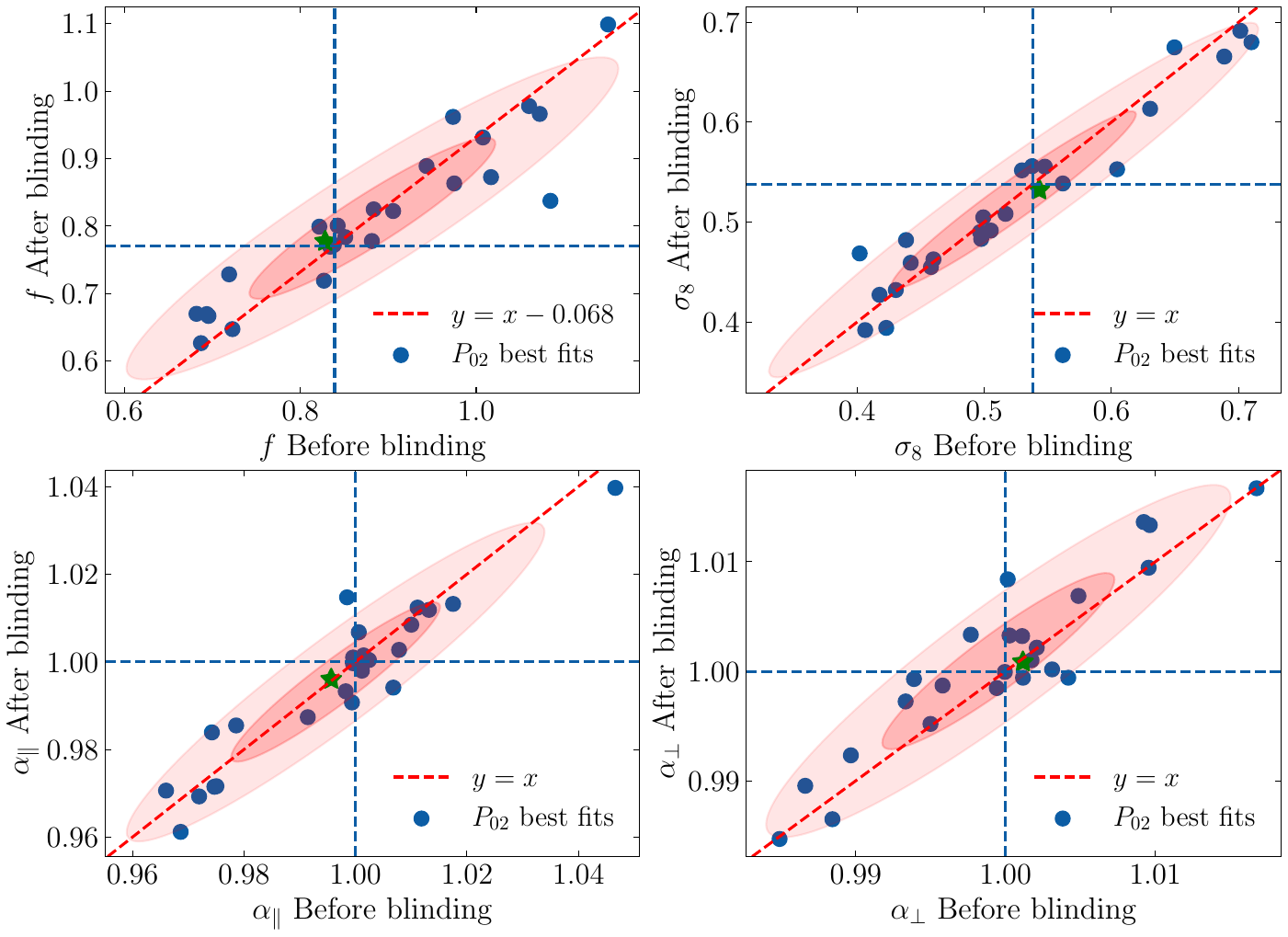}
\caption{Recovered cosmological parameters $\{f,\sigma_8,\alpha_\parallel,\alpha_\bot\}$  before and after blinding for the power spectrum data-vector $P_{02}=\{P_0,P_2\}$ from 25 \textsc{AbacusSummit} LRG cubic boxes. The scatter shows the relationship between the maximum likelihood parameters before ($x$-axis) and after ($y$-axis) applying RSD-type blinding. The blue dashed lines indicate the underlying/expected value for each parameter, whilst the red dashed lines indicate the expected relationship for the parameters between the original and blinded realizations. The star points are obtained using the mean of the power spectra of the 25 boxes as data-vector, and the red ellipses represent the 1 and 2$\sigma$ regions of the scatter of the 25 realizations, computed with the eigenvalues and eigenvectors of the covariance of the scatter. }
\label{fig: P02_f73blind_unblind_correlation}
\end{figure}

\begin{figure}[ht!]
\centering 
\includegraphics[width = \textwidth]{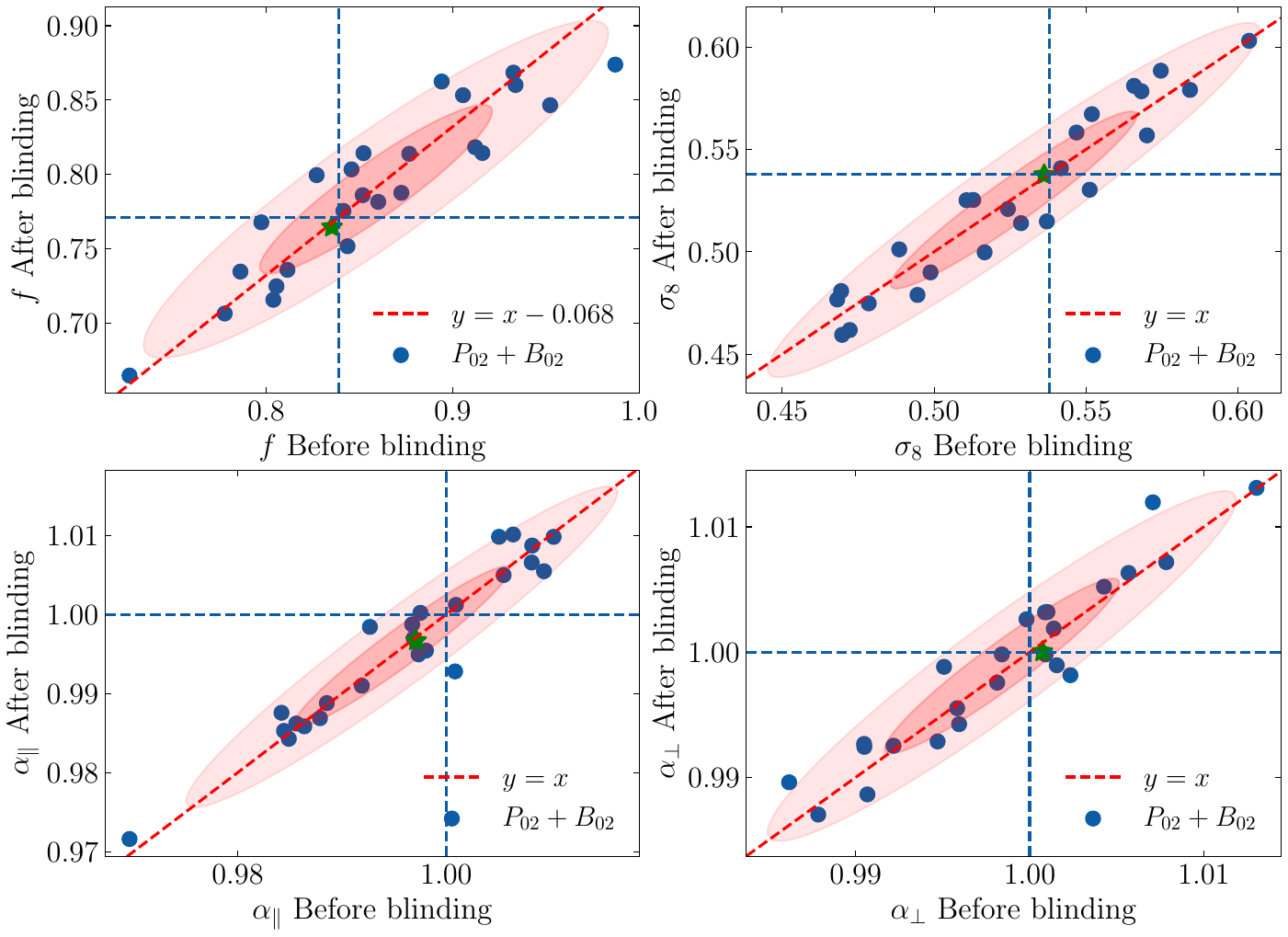}
\caption{Analogous to Figure \ref{fig: P02_f73blind_unblind_correlation} but for the data-vector $P_{02}$+$B_{02}=\{P_0,P_2,B_0,B_{200},B_{020}\}$ as indicated in Section \ref{sec: 3}. Original and blinded recovered parameters follow the expected distribution.  The inclusion of the bispectrum significantly reduces the scatter and thus the errors.}
\label{fig: B02k11_f73blind_unblind_correlation}
\end{figure}

The results are displayed in Figure \ref{fig: P02_f73blind_unblind_correlation}, where we plot the scatter among the 25 boxes of the maximum likelihood of the cosmological parameters $\{f,\sigma_8,\alpha_\parallel,\alpha_\bot\}$ before and after blinding. In this figure, each of the filled blue circles corresponds to one of the 25 boxes, and the coordinates represent the maximum likelihood parameter values before ($x$ axis) and after ($y$ axis) blinding. The green stars correspond to the maximum likelihood values for the mean signal from the 25 boxes. The expectation is that symbols should scatter  around the line of equation $y=x+\Delta\Omega$, where $\Delta\Omega$ is the  (blinding) shift in the given parameter.\footnote{Note that $\Delta\Omega=0$ for parameters  not expected to be affected by  blinding. In this test for RSD blinding in boxes, the parameters with $\Delta\Omega=0$ are $\{\sigma_8,\alpha_\parallel,\alpha_\bot\}$.} This corresponds to the red dashed line. The true underlying (and expected) values of the parameters before (and after) blinding are indicated by the dashed vertical and horizontal blue lines. This preliminary step shows that we can replicate the results obtained in \cite{brieden2020blind}. As expected, if only $P_{02}$ is used, the $f$--$\sigma_8$ degeneracy is only weakly broken by the non-linear terms of the power spectrum monopole and quadrupole. 
This degeneracy is not fully disentangled unless the bispectrum monopole and quadrupoles are included in the data-vector \cite{novell2023geofpt,Gualdi:2020aniso}.

\begin{figure}[ht!]
\centering 
\includegraphics[width = \textwidth]{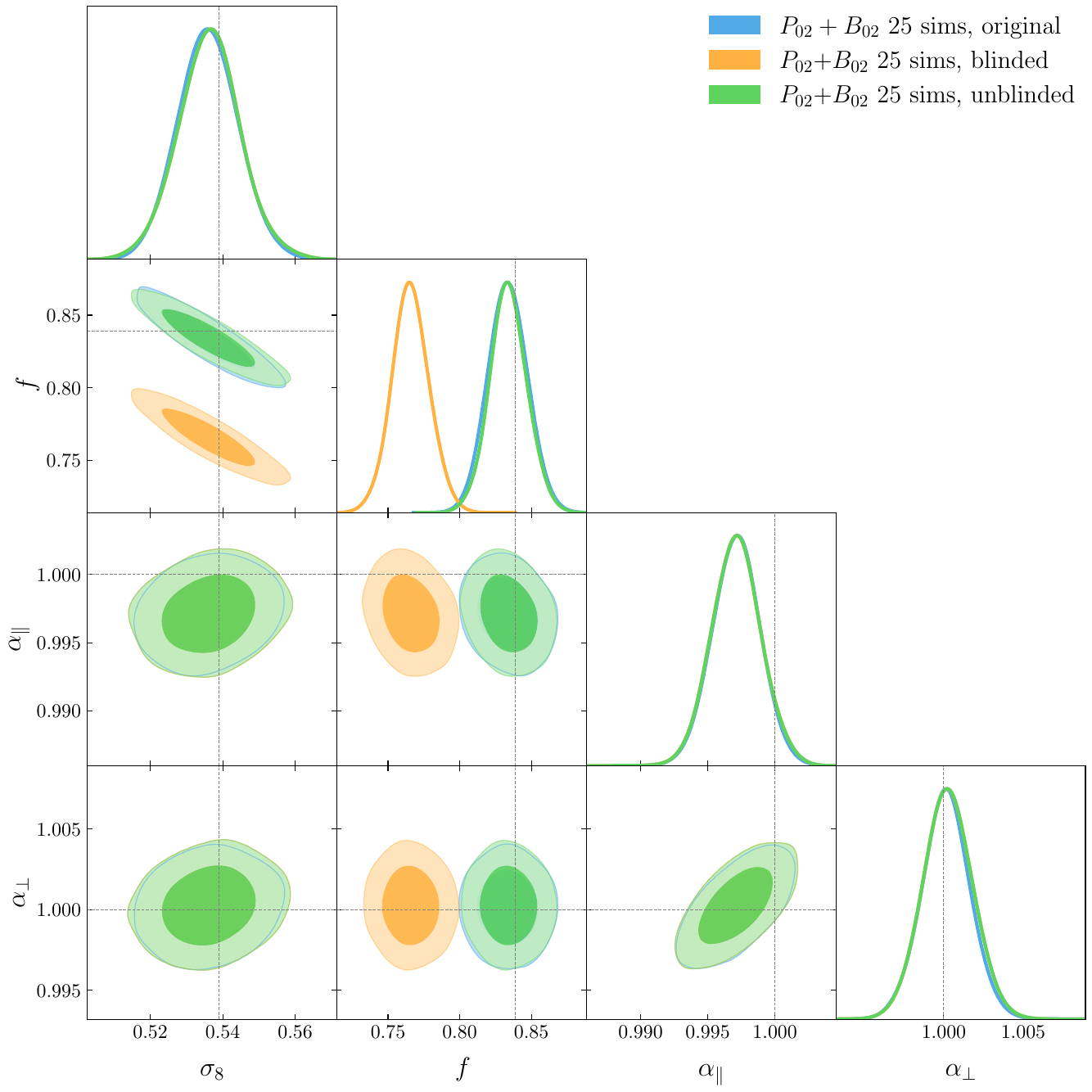}
\caption{Cubic boxes, RSD blinding only. Recovered 1-2D posteriors for the cosmological parameters $\{f,\sigma_8,\alpha_\parallel,\alpha_\bot\}$ using the mean $P_{02}$+$B_{02}$ from the 25 \textsc{AbacusSummit} LRG boxes as the data-vector. The statistical errors correspond to the full volume of the 25 boxes, $V_{25}=25\times2^3\,(\textrm{Gpc}\,h^{-1})^3$. The blue and orange contours are, respectively, the results from the original and RSD-blinded simulations, while the green \textit{unblinded} contours are obtained by analytically shifting back the posteriors by the expected $\Delta f$. Even though the effective volume ($\sim 160\,(\textrm{Gpc}\,h^{-1})^3$ \cite{tegmark1997measuring}) is approximately an order of magnitude larger than the expected DESI volume for LRG's, the unblinded case is almost identical to the original, while recovering the expected values at $\sim1\sigma$ level, thus providing validation of the use of both the GEO-FPT bispectrum model and the RSD part of the blinding scheme.}
\label{fig: Abacus_blind_posteriors}
\end{figure}

The corresponding results of the $P_{02}$+$B_{02}$ joint analysis are shown in Figure \ref{fig: B02k11_f73blind_unblind_correlation} using the same conventions.
As with the case of the power spectrum, we find excellent agreement between original and blinded results, given the expected shift in $f$. Note the different scales in the figure: the errorbars on the parameters are significantly reduced by including the bispectrum monopole and quadrupoles in the analysis.

The cosmological constraints obtained with the mean signal of the 25 simulations and the corresponding covariance (i.e., for a physical volume of $V_{25}=25\times2^3\,(\textrm{Gpc}\,h^{-1})^3$) are shown in Figure \ref{fig: Abacus_blind_posteriors}.
The original (blue), blinded (orange) and unblinded (green) posteriors for all the possible pairs of the $\{f,\sigma_8,\alpha_\parallel,\alpha_\bot\}$ parameters are shown. By ``unblinded''  we refer to the posteriors of the blinded case numerically shifted back by the designed shift i.e. by  $-\Delta f=0.068$.
As we mentioned in Section \ref{sec: 3}, the shift in $f$ produced by the blinding procedure is equivalent to $\sim3\sigma$ of the parameter error in the $P_{02}$+$B_{02}$ analysis in the joint volume covered by the 25 simulations.

We can see that the original analysis, including the bispectrum multipoles, recovers the underlying cosmological parameters within 1$\sigma$, even for a volume (and thus statistical precision)  that far exceeds the total DESI volume.
The overlap of the original (blue) and unblinded (green) posterior shows that the RSD blinding component shifts the recovered posteriors exactly as predicted.

\subsection{Cutsky mocks}
\label{sec: cutsky}
The AP component of the blinding scheme, by acting simply on the fiducial cosmology adopted in the redshift-distance relation, by design shifts the dilation parameters exactly as modeled in all the established BAO analyses, e.g. \cite{eisenstein2005detection,bautista2021completed}.

\begin{figure}[ht!]
\centering 
\includegraphics[width = \textwidth]{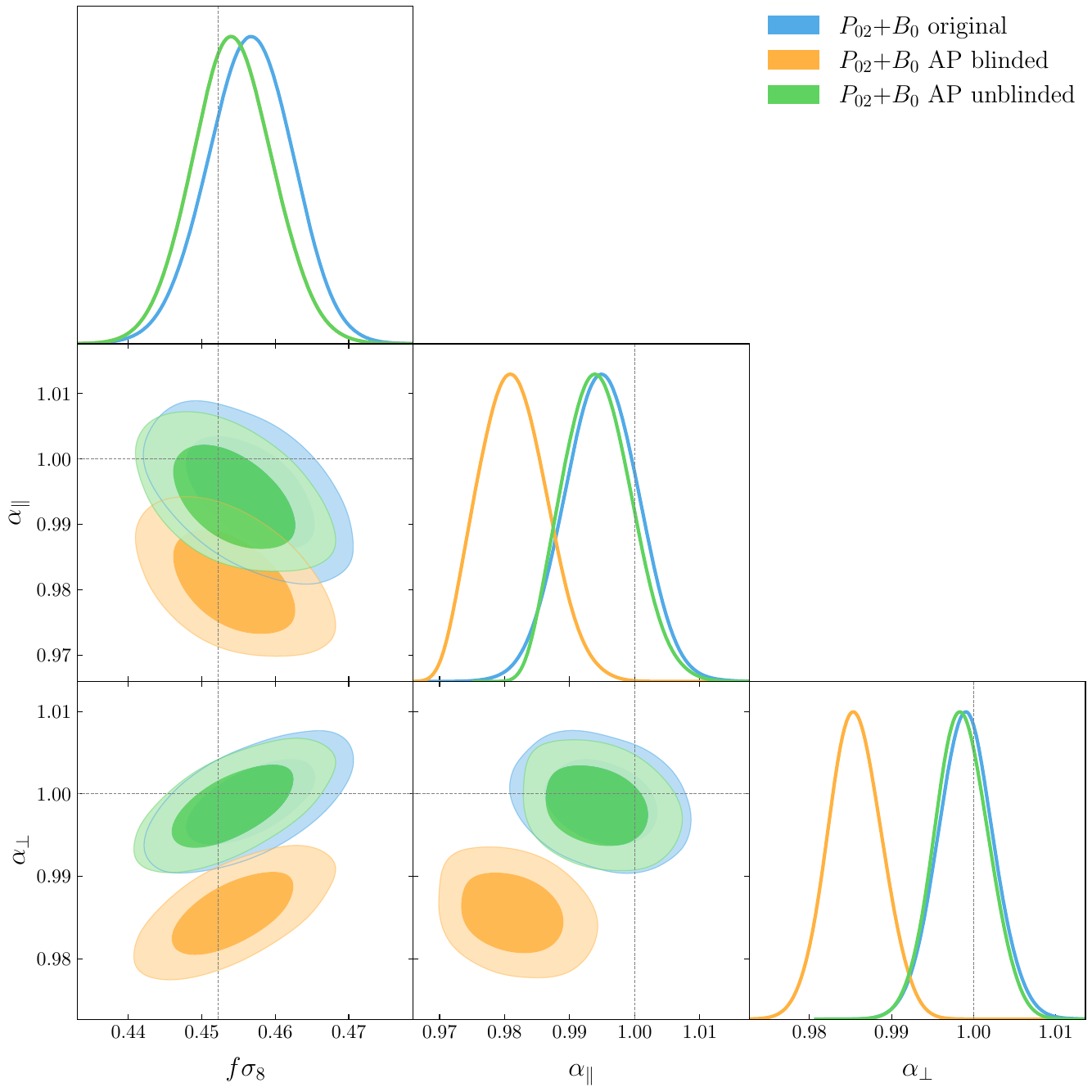}
\caption{Analogous to Figure \ref{fig: Abacus_blind_posteriors},  using the same conventions but for cutsky boxes, and AP blinding only. Recovered 1-2D posteriors for the cosmological parameters $\{f\sigma_8,\alpha_\parallel,\alpha_\bot\}$ using the mean $P_{02}$+$B_{0}$ from the 25 \textsc{AbacusSummit} LRG cutsky mocks at $z=0.8$. The single realization covariance is rescaled by 25 so that it matches the total volume of the 25 mocks, $V^\textrm{eff}_{\rm 25cutsky}\sim 40 \,(\textrm{Gpc}\,h^{-1})^3$. We model the effect of the window function according to the approximation in Section \ref{sec: 3}. Similarly to Figure \ref{fig: Abacus_blind_posteriors}, the AP blinding operation shifts the posteriors exactly as expected. In this case, the blinding procedure shifts the parameters as $\Delta\alpha_\parallel=\Delta\alpha_\bot=-0.013$, which corresponds to $1\sigma$ of the parameters errorbars. In the 1D distribution for $f\sigma_8$ the orange curve is identical to the green one and thus invisible.}
\label{fig: Abacus_cutsky_p02b0blindingAP}
\end{figure}

Nevertheless, for the sake of systematically validating the blinding pipeline, we first test the behaviour of the AP blinding in the presence of a survey window, in isolation of the RSD part, which corresponds to our second step as mentioned in Section \ref{sec: theory_blinding}. In Figure \ref{fig: Abacus_cutsky_p02b0blindingAP} we show the posterior distribution for the main cosmological parameters of our analysis, $\{f\sigma_8,\alpha_\parallel,\alpha_\bot\}$ for both the unblinded and AP blinded cutsky mocks, together with the shifted (``unblinded'', in green) posteriors. The statistical errors correspond to the cosmological volume spanned by the 25 individual mocks, which is $\sim 40\, (\textrm{Gpc}\,h^{-1})^3$ as explained in Section \ref{sec: 3}. 

In this application to cutsky mocks, we report constraints on the product $f\sigma_8$ instead of the separate parameters $f$ and $\sigma_8$ since the bispectrum monopole cannot fully resolve the well-known $f$--$\sigma_8$ degeneracy \cite{novell2023geofpt}. The inclusion of the bispectrum quadrupoles (once a suitable modeling of the window becomes available) is expected to resolve this. However, at this time, this is beyond the scope of this work, centered on validating the blinding scheme.

The figure shows that the posterior distributions for the parameters of interest are all within 1$\sigma$ of the true values, which is a validation of both the power spectrum and bispectrum models and bias expansions, together with the window approximation we use for $B_0$. The close similarity of the blue (before blinding) and green (after blinding,  but numerically shifted back) posterior contours indicates that for the adopted magnitude of the blinding shift of $\alpha_\parallel=\Delta\alpha_\bot=-0.013$ (corresponding to $1$--$2\sigma$ of our errorbars for the total of 25 mocks), the blinding procedure performs as expected.

\begin{figure}[ht!]
\centering 
\includegraphics[width = \textwidth]{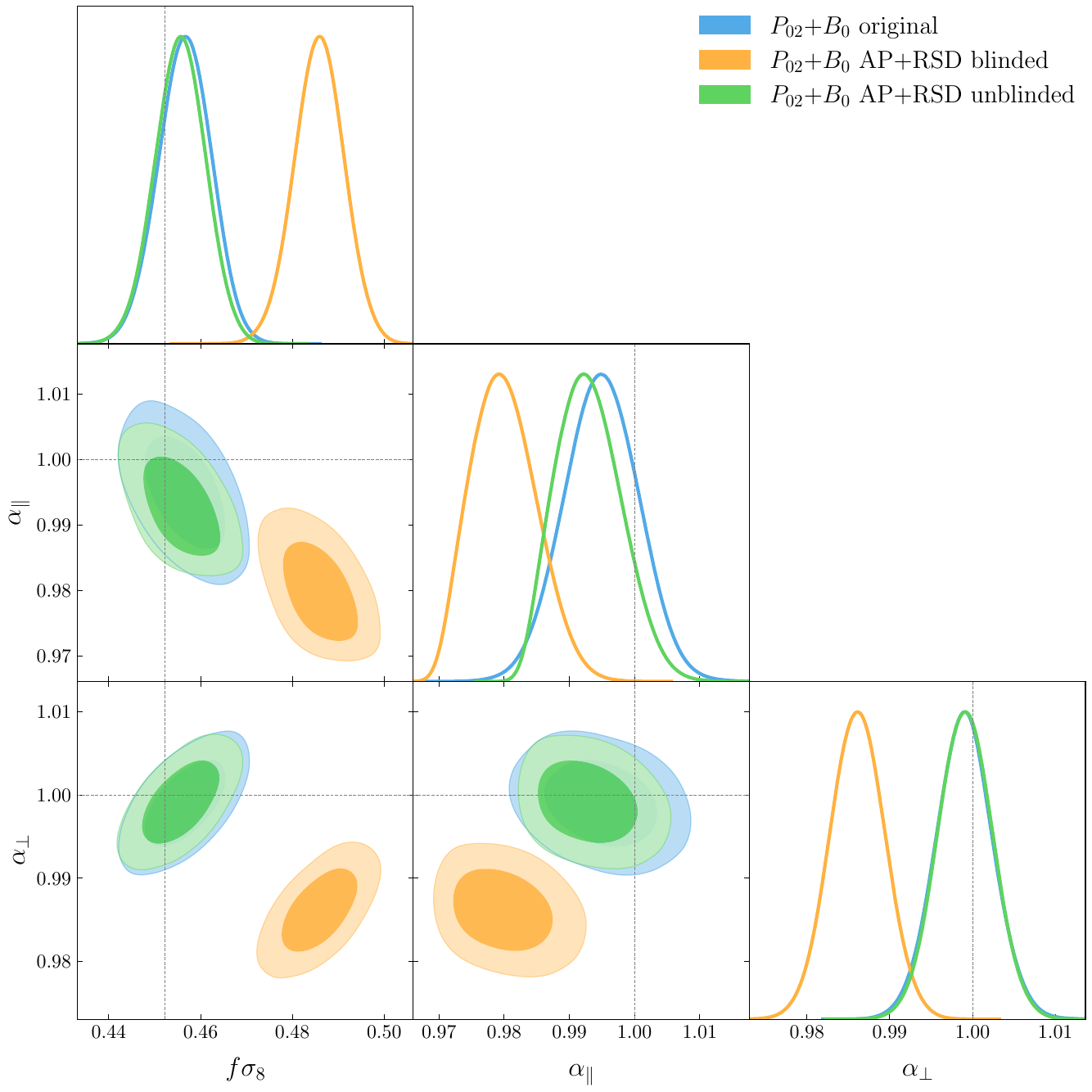}
\caption{Analogous plot as Figure \ref{fig: Abacus_cutsky_p02b0blindingAP}, in this case performing the full blinding procedure, which also includes the RSD shift as described in Section \ref{sec: theory_blinding}. Both AP and RSD parts of the blinding are seen to perform exactly as expected. }
\label{fig: Abacus_cutsky_p02b0blindingAPRSD}
\end{figure}

Finally,  Figure \ref{fig: Abacus_cutsky_p02b0blindingAPRSD} shows the results of the test of the full blinding pipeline---AP and RSD---applied to the cutsky mocks, with shifts of $\Delta f=0.06,\,\Delta \alpha_\parallel=\Delta\alpha_\bot=-0.013$. 
The $\alpha_\parallel,\alpha_\bot$ parameters are (correctly) shifted as in the case without the RSD part of the blinding:   even with the addition of the bispectrum, and in the presence of a sky cut, the two parts of the blinding do not interfere significantly.
As for the redshift-space distortions parametrization $f\sigma_8$, the RSD part of the blinding we applied is designed to shift $f$ while not affecting $\sigma_8$, and the recovered posterior distribution for $f\sigma_8$ is exactly as predicted. 

The  original and shifted (unblinded) posteriors in Figures \ref{fig: Abacus_cutsky_p02b0blindingAP} and \ref{fig: Abacus_cutsky_p02b0blindingAPRSD} indicates that estimated blind parameter constraints are slightly tighter than the original ones.  This is not unexpected, as for this application we do not transform the covariance matrix under the blinding operation. The clustering properties of the EZmocks used to compute the covariance are designed to match those of the (original) before blinding catalogs. Blinding changes slightly the clustering properties making the covariance matrix slightly mis-calibrated. While this effect can be corrected---albeit in a somewhat time-consuming way, by applying the blinding transformation to the 1000 realizations of EZmocks---for this application it is small enough that can be left uncorrected.

\section{Conclusions}
\label{sec: conclusions}
In the era of large stage IV galaxy surveys, blind analyses have become essential to safeguard against confirmation bias in cosmology, where researchers may unconsciously influence their analyses to conform to their prior beliefs. By masking certain aspects of the data or results until the analysis is complete, blinding adds to the integrity of findings from present and upcoming galaxy surveys \cite{brieden2020blind,muir2020blinding}.

The blinding technique of \cite{brieden2020blind} has been validated for two-point statistics and currently adopted by and implemented in the official DESI pipeline. This blinding scheme modifies the distance-redshift relation and thus the tracer's redshifts in two complementary parts. The first part, named AP blinding in this work, is equivalent to modifying the cosmology used in transforming redshifts into distances, by changing the default $\Lambda$CDM values of the $w_0,w_a$ parameters in the dark energy equation of state parameterization given by  $w(z)=w_0+(1-a(z))w_a$. This generates a (theoretically) predictable shift in the recovered values of $\alpha_\parallel,\alpha_\bot$. The second part of blinding, the RSD blinding, combines the observed density field with a suitably reconstructed density field in order to modify the effective value of the linear growth rate parameter $f$.

In this paper 
we have explored the effectiveness and performance of the catalog-level blinding technique proposed by \cite{brieden2020blind} within DESI-like galaxy spectroscopic surveys, for a data-vector which combines the power spectrum and bispectrum summary statistics {, using a fixed template approach. In the case of analyses using instead a full modelling approach (e.g. \cite{carrasco2012effective,nishimichi2020blinded,chudaykin2020nonlinear,colas2020efficient,ivanov2020cosmological,d2022boss}) the blinding scheme would need to be validated in a separate work.}

We have confirmed that such a blinding scheme performs as expected when including the bispectrum data-vector,  in the three sequential tests below:
\begin{enumerate}
    \item  RSD-only blinding on cubic boxes. For a blinding shift of $\Delta f=-0.068$ (which corresponds to $\sim3\sigma$ of the recovered errorbars\footnote{In all cases, both cubic and cutsky mocks, the total volume of both the signal and covariance corresponds to the sum of the 25 available mocks. This yields an effective volume of $\sim 160\,(\textrm{Gpc}\,h^{-1})^3$ in the case of cubic mocks and of $\sim 40\,(\textrm{Gpc}\,h^{-1})^3$ in the cutsky mocks.} in $f$) and with a data-vector consisting of power spectrum and bispectrum monopole and quadrupoles $\{P_0,P_2,B_0,B_{200},B_{020}\}$, we found perfect agreement between the predicted blinded and recovered blinded parameters.
    \item  AP-only blinding on cutsky mocks, with a data-vector composed of power spectrum multipoles and bispectrum monopole  $\{P_0,P_2,B_0\}$. The modeling of the window function on the bispectrum monopole follows the approximation proposed by \cite{Gil-Marin:2014biasgravity, Gil-Marin:2016sdss}.  As expected, given that this component is directly modifying the cosmology---which the $\alpha_\parallel,\alpha_\bot$ parameters naturally quantify---the parameters recovered in the blinded mocks closely follow the predictions. In this case, we shifted the $w_0,w_a$ parameters such that the expected shifts are $\Delta\alpha_\parallel=\Delta\alpha_\bot=-0.013$, about 1--2$\sigma$ of errorbars obtained in $\alpha_\parallel,\alpha_\bot$. 
    
    \item Full blinding pipeline, including both AP and RSD parts on cutsky mocks for $\{P_0,P_2,B_0\}$ data vector. The AP part is implemented as above, and the RSD part yields a theoretical shift of $\Delta f=0.060$.  The AP part behaves exactly as expected. For this data vector, the RSD part can only be reliably tested by considering the parameter combination $f\sigma_8$: the addition of the bispectrum monopole is not sufficient to break the well-known degeneracy between these two parameters (see \cite{novell2023geofpt} for a  recent discussion). 
\end{enumerate}

We conclude that these results provide sufficient validation to the blinding pipeline for present-day analyses involving the galaxy bispectrum.
To date, there is no sufficiently precise modeling of the survey window function on the bispectrum multipoles. Should one become available, steps 2 and 3 above could be extended to the full data vector including bispectrum multipoles. We leave this to future work.

The work  presented in this paper represents the first step to validate our pipeline for the joint power spectrum and bispectrum analysis of the DESI survey. Forthcoming work will include  improvements of the survey window approximation for the bispectrum and the treatment of the imaging and systematic weights on the bispectrum.

\section*{Data Availability}
All data from the tables figures are available in machine-readable format at
\href{https://zenodo.org/uploads/11984896?token=eyJhbGciOiJIUzUxMiJ9.eyJpZCI6IjhkYmUyODdkLWIxZTQtNGUxNS05YzExLWJhNzk4MWQxMjllOCIsImRhdGEiOnt9LCJyYW5kb20iOiJhYjQxOGRjYzc2NTIwZGRlOTlkMDAxNTVkNTU5ZDc1YyJ9.FTTFsCqQ6v_hDLvHZ7EVyXhLmgJZcrNEZYMPFlnWE1gFRkXIG1KJX-9ak67RIfpXWJFg6byDC13qc9yVnPjL7A}{10.5281/zenodo.11984896} in compliance with the DESI
data management plan.

\section*{Acknowledgements}

SNM, HGM and LV thank Santiago Ávila, Mike Shengbo Wang and Benjamin Weaver for the helpful discussion and suggestions.
SNM acknowledges funding from the official doctoral program of the University of Barcelona for the development of a research project under the PREDOCS-UB grant.
HGM acknowledges support through the program Ram\'on y Cajal (RYC-2021-034104) of the Spanish Ministry of Science and Innovation. 
LV and HGM acknowledge the support of the European Union’s Horizon 2020 research and innovation program ERC (BePreSySe, grant agreement 725327).

Funding for this work was partially provided by the Spanish MINECO under project PGC2018-098866-B-I00MCIN/AEI/10.13039/501100011033 y FEDER ``Una manera de hacer Europa'', and the ``Center of Excellence Maria de Maeztu 2020-2023'' award to the ICCUB (CEX2019-000918-M funded by MCIN/AEI/10.13039/501100011033).

This material is based upon work supported by the U.S. Department of Energy (DOE), Office of Science, Office of High-Energy Physics, under Contract No. DE–AC02–05CH11231, and by the National Energy Research Scientific Computing Center, a DOE Office of Science User Facility under the same contract. Additional support for DESI was provided by the U.S. National Science Foundation (NSF), Division of Astronomical Sciences under Contract No. AST-0950945 to the NSF’s National Optical-Infrared Astronomy Research Laboratory; the Science and Technology Facilities Council of the United Kingdom; the Gordon and Betty Moore Foundation; the Heising-Simons Foundation; the French Alternative Energies and Atomic Energy Commission (CEA); the National Council of Humanities, Science and Technology of Mexico (CONAHCYT); the Ministry of Science and Innovation of Spain (MICINN), and by the DESI Member Institutions: \url{https://www.desi.lbl.gov/collaborating-institutions}. Any opinions, findings, and conclusions or recommendations expressed in this material are those of the author(s) and do not necessarily reflect the views of the U. S. National Science Foundation, the U. S. Department of Energy, or any of the listed funding agencies.

The authors are honored to be permitted to conduct scientific research on Iolkam Du’ag (Kitt Peak), a mountain with particular significance to the Tohono O’odham Nation.

This work has made use of the following publicly available codes: \href{https://github.com/serginovell/geo-fpt}{\textsc{GEO-FPT}} \cite{novell2023geofpt}, \href{https://github.com/hectorgil/Rustico}{\textsc{Rustico}} \cite{HGMeboss}, \href{https://github.com/hectorgil/Brass}{\textsc{Brass}} \cite{HGMeboss}, \href{https://emcee.readthedocs.io/en/stable/index.html}{\textsc{Emcee}} \cite{Foreman_Mackey_2013}, \href{https://www.gnu.org/software/gsl/}{GSL} \cite{gsl}, \href{https://scipy.org/}{\textsc{SciPy}} \cite{scipy}, \href{https://numpy.org/}{\textsc{NumPy}} \cite{numpy}, \href{https://getdist.readthedocs.io/en/latest/}{\textsc{GetDist}} \cite{getdist}, \href{https://www.astropy.org}{\textsc{Astropy}} \cite{astropy}, \href{https://matplotlib.org}{\textsc{Matplotlib}} \cite{matplotlib}. We are grateful to the developers who made these codes public.

\appendix
\section{Additional blinding considerations}
\label{sec: app}
\begin{figure}[ht!]
\centering 
\includegraphics[width = \textwidth]{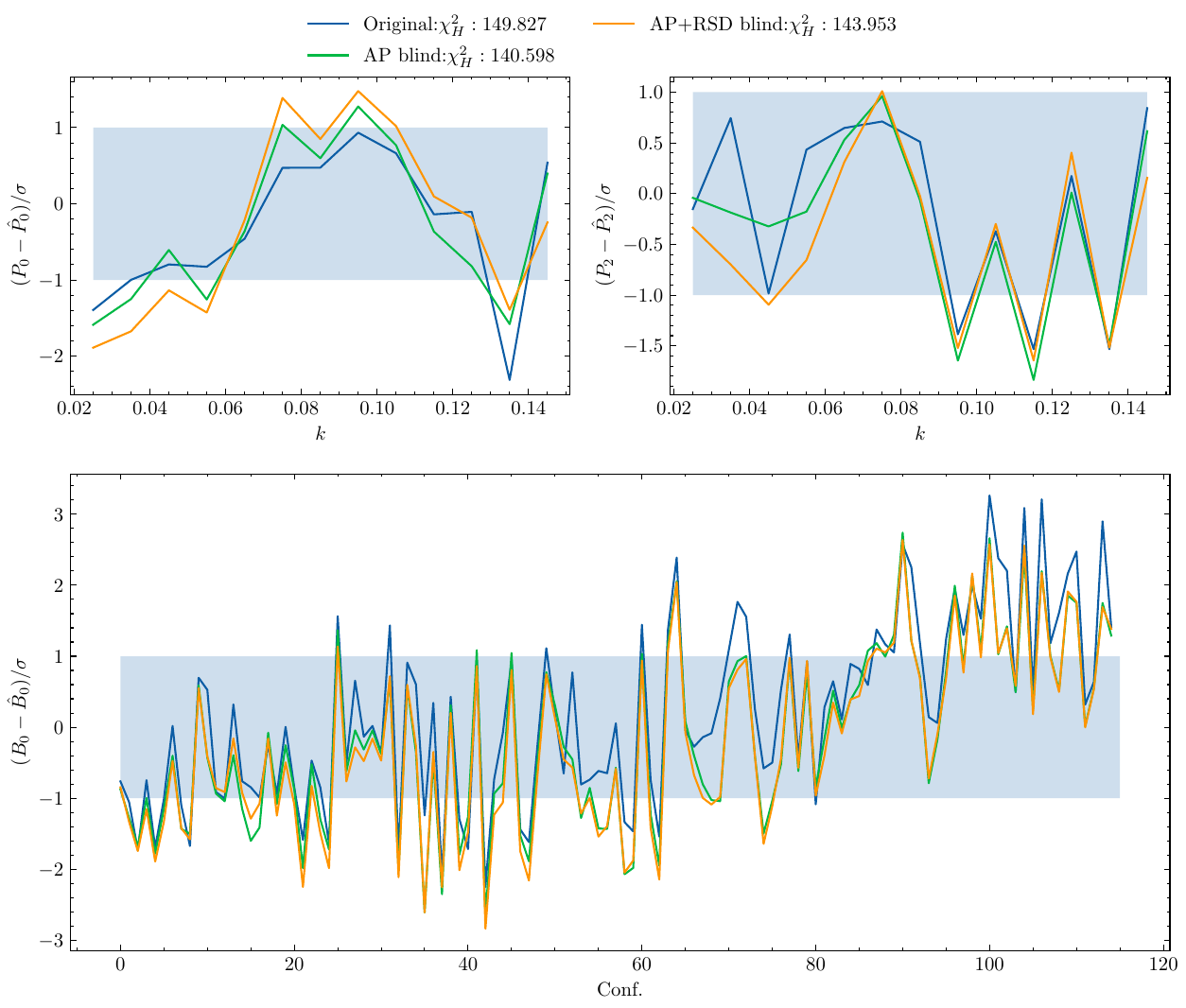}
\caption{Best fits for the cutsky mocks for the three components of the data-vector, $\{P_0,P_2,B_0\}$, in the cases where the mocks were not blinded, AP-blinded, and fully blinded (AP+RSD). The quantity that is shown is the discrepancy between the best-fitting model and the data (eg. $P_0-\hat{P}_0$) divided by the corresponding error $\sigma$. The shaded regions delimit the 1$\sigma$ region. The $x$-axis in the lower plot refers to the triangle configuration index.}
\label{fig: ratios_blinding}
\end{figure}

We show in Figure \ref{fig: ratios_blinding} the best fitting theoretical model for the joint data-vector $\{P_0,P_2,B_0\}$ of the cutsky mocks in the original and blinded cases (both AP-only and AP+RSD). There is no significant difference in the $\chi_H^2$ values (where the $H$ subscript stands for Hartlap-corrected covariance matrix) when the data-vector is blinded. Furthermore, the values for $\chi^2_H$ are close to the number of degrees of freedom, which is equal to the number of elements of the full data-vector (141) minus the number of free parameters (9). Even in this case, where the errorbars correspond to an effective volume of $\sim40\,(\textrm{Gpc}\,h^{-1})^3$, the values for $\chi_H^2$ are all within $\sim10\%$ of the number of degrees of freedom. This (along with the results shown in the main text) is an indication that the adopted models for the power spectrum and bispectrum do not show evidence for large systematic errors given the expected statistical error of present and upcoming surveys. 

\section{Theoretical modelling of power spectrum and bispectrum}
\label{app: theory}
In this section we describe the theoretical formalism that we use for the power spectrum and bispectrum, both rooted in Standard Perturbation Theory (SPT hereafter). 

The redshift space galaxy power spectrum, $P_g$, is computed from the non-linear matter power spectrum, $P_{g,\delta\delta}$, the density-velocity, $P_{g,\delta\theta}$, and velocity-velocity, $P_{g,\theta\theta}$, power spectra, according to the TNS model \cite{Taruya_2010,Nishimichi_2011},
\begin{align}
\label{eq: Pnl_redshift}
     P_g(k,\mu)&=D_\textrm{FoG}^P(k,\mu,\sigma_P)\big[P_{g,\delta\delta}(k)+2f\mu^2P_{g,\delta\theta}(k)+f^2\mu^4P_{\theta\theta}\nonumber\\
     &+b_1^2A^\textrm{TNS}(k,\mu,f/b_1)+b_1^4B^\textrm{TNS}(k,\mu,f/b_1)\big],
\end{align}
where $f$ denotes the  logarithmic growth rate of perturbations, and $d \ln \delta /d\ln a$ and  $P_{g,\delta\delta}, P_{g,\delta\theta}$ are computed as in \cite{beutler2014clustering}. In doing so, we are using the bias expansion $\{b_1,b_2,b_{s^2},b_\textrm{3nl}\}$ and assuming the Lagrangian local bias approximation. Additionally, the functions $A^\textrm{TNS},B^\textrm{TNS}$ are  defined in \cite{Taruya_2010}, $\mu$ is the cosine of the angle of $k$ with the line of sight, and $D_\textrm{FoG}^P$ is a damping factor that accounts for the Fingers-of-God (FoG) effect of  redshift space distortions \cite{Jackson_1972}. We model the FoG damping factor for the bispectrum as  
\begin{equation}
    D_\textrm{FoG}^P(k,\mu,\sigma_\textrm{FoG}^P)=\frac{1}{\left(1+k^2\mu^2\sigma_P^2/2\right)^2},
\end{equation}
where $\sigma_P$ is a free parameter to be constrained by the data.

The SPT redshift space bispectrum can then be written in the following way at tree-level order:
\begin{equation}
B^{\rm SPT}(\textbf{k}_1,\textbf{k}_2,\textbf{k}_3)=D_\textrm{FoG}^B(\textbf{k}_1,\textbf{k}_2,\textbf{k}_3)\left[2Z_1^\textrm{SPT}(\textbf{k}_1)Z_1^\textrm{SPT}(\textbf{k}_2)Z_2^\textrm{SPT}(\textbf{k}_1,\textbf{k}_2)P^L(k_1)P^L(k_2) + \textrm{2perm.}\right],
\label{eq:bispredshiftsp}
\end{equation}
where the kernels $Z_1^\textrm{SPT},Z_2^\textrm{SPT}$ are computed as
\begin{align}
    Z_1^\textrm{SPT}(\textbf{k})&=b_1+f\mu^2,\nonumber\\
    Z_2^\textrm{SPT}(\textbf{k}_1,\textbf{k}_2)&=b_1F_2^\textrm{SPT}(\textbf{k}_1,\textbf{k}_2)+f\mu_{12}^2G_2^\textrm{SPT}(\textbf{k}_1,\textbf{k}_2)+\frac{b_1f}{2}\left(\mu_1^2+\mu_2^2+\mu_1\mu_2\left(\frac{k_1}{k_2}+\frac{k_2}{k_1}\right)\right)\nonumber\\
    &+f^2\mu_1\mu_2\left(\mu_1\mu_2+\frac{1}{2}\left(\mu_1^2\frac{k_1}{k_2}+\mu_2^2\frac{k_2}{k_1}\right)\right)+\frac{1}{2}\left(b_2+b_{s^2}S_2^\textrm{SPT}(\textbf{k}_1,\textbf{k}_2)\right),\label{eq: z1z2}
\end{align}
while $\mu_{ij}\equiv(k_i\mu_i+k_j\mu_j)/|\textbf{k}_i+\textbf{k}_j|$. The $G_2^\textrm{SPT}$ and $S_2^\textrm{SPT}$ kernels in SPT are 
\begin{align}
    G_2^\textrm{SPT}(\textbf{k}_1,\textbf{k}_2)&=\frac{3}{7}+\frac{1}{2}\cos(\theta_{12})\left(\frac{k_1}{k_2}+\frac{k_2}{k_1}\right)+\frac{4}{7}\cos^2(\theta_{12}),\\
    S_2^\textrm{SPT}(\textbf{k}_1,\textbf{k}_2)&=\cos(\theta_{12})^2-\frac{1}{3}.
\end{align}

In this work, we use the GEO-FPT bispectrum model \cite{novell2023geofpt}, which introduces a modification to the $Z_2$ kernel, as follows:
\begin{equation}\label{eq: Z2_geo}
    Z_2^{\textrm{GEO}} = Z_2^\textrm{SPT}\times\Big[f_1+f_2\frac{\cos(\theta_\textrm{med})}{\cos(\theta_\textrm{max})}+f_3\frac{\cos(\theta_\textrm{min})}{\cos(\theta_\textrm{max})}+f_4\frac{A}{A_\textrm{norm}}+f_5\frac{A^2}{A_\textrm{norm}^2}\Big].
\end{equation}
where the coefficients $f_1,...,f_5$  were calibrated on N-body simulations \cite{novell2023geofpt}.

For the bispectrum, we parametrize the FoG damping factor as \cite{Scoccimarro:1999ed,Verde1998}
\begin{equation}
\label{eq:fog_lorentz}
    D_\textrm{FoG}^B(\textbf{k}_1,\textbf{k}_2,\textbf{k}_3)=(1+\left[k_1^2\mu_1^2+k_2^2\mu_2^2+k_3^2\mu_3^2\right]^2\sigma_{B}^4/2)^{-2},
\end{equation}
again with $\sigma_B$ being a free parameter.

We model the deviations from Poissonian shot-noise with the parameters $A_\textrm{P},A_\textrm{B}$, which modify the Poisson prediction as in \cite{Gil-Marin:2014theory,gil-marin_clustering_2017}:
\begin{align}
    P_{\textrm{noise}}&=(1-\frac{A_\textrm{P}}{\alpha_\parallel \alpha_\bot^2})P_\textrm{Poisson},\\
    B_{\textrm{noise}}(k_1,k_2,k_3)&=(1-\frac{A_\textrm{B}}{\alpha_\parallel^2\alpha_\bot^4})B_\textrm{Poisson}(k_1,k_2,k_3).
\end{align}

The power spectrum and bispectrum redshift space multipoles are then obtained by integrating the expansion of the power spectrum and bispectrum dependence on the angle with respect to the line of sight in terms of Legendre polynomials $\mathcal{L}_i$, so that

\begin{align}
    P^{(\ell)}(k)&=\frac{2\ell+1}{2\alpha_\parallel\alpha_\bot^2}\int_{-1}^1d\mu P(k,\mu)\mathcal{L}_{\ell}(\mu),\label{eq: pmulti}\\
    B^{(\ell_i)}(\textbf{k}_1,\textbf{k}_2,\textbf{k}_3)&=\frac{2\ell+1}{4\pi\alpha_\parallel^2\alpha_\bot^4}\int_{-1}^1d\mu_1\int_0^{2\pi}d\phi B(\textbf{k}_1,\textbf{k}_2,\textbf{k}_3)\mathcal{L}_{\ell}(\mu_i),
\end{align}
Here $\phi$ is defined as the angle fulfilling $\mu_2=\mu_1\cos\theta_{12}-\sqrt{(1-\mu_1^2)(1-\cos\theta_{12}^2)}\cos\phi$, and $\ell_i$ refers to the multipole of order $\ell$ ($\ell=0,2$ corresponding respectively to the monopole and quadrupole), and the $i$ index denotes which multipole (for instance $\ell=2,$ $\ell_1$ being the quadrupole $(200)$). The power spectrum multipole expansion of \ref{eq: pmulti} was proposed in \cite{hamilton1992measuring,cole1994fourier}, while the bispectrum expansion and choice of variables was first used in \cite{Scoccimarro:1999ed}.

\bibliographystyle{ieeetr}
\bibliography{main}

\begin{thebibliography}{10}

\bibitem{benson2014spt}
B.~A. Benson, P.~Ade, Z.~Ahmed, S.~Allen, K.~Arnold, J.~Austermann, A.~Bender,
  L.~Bleem, J.~Carlstrom, C.~Chang, {\em et~al.}, ``Spt-3g: a next-generation
  cosmic microwave background polarization experiment on the south pole
  telescope,'' in {\em Millimeter, Submillimeter, and Far-Infrared Detectors
  and Instrumentation for Astronomy VII}, vol.~9153, pp.~552--572, SPIE, 2014.

\bibitem{aiola2020atacama}
S.~Aiola, E.~Calabrese, L.~Maurin, S.~Naess, B.~L. Schmitt, M.~H. Abitbol,
  G.~E. Addison, P.~A. Ade, D.~Alonso, M.~Amiri, {\em et~al.}, ``The atacama
  cosmology telescope: Dr4 maps and cosmological parameters,'' {\em Journal of
  Cosmology and Astroparticle Physics}, vol.~2020, no.~12, p.~047, 2020.

\bibitem{galitzki2018simons}
N.~Galitzki, A.~Ali, K.~S. Arnold, P.~C. Ashton, J.~E. Austermann,
  C.~Baccigalupi, T.~Baildon, D.~Barron, J.~A. Beall, S.~Beckman, {\em et~al.},
  ``The simons observatory: instrument overview,'' in {\em Millimeter,
  Submillimeter, and Far-Infrared Detectors and Instrumentation for Astronomy
  IX}, vol.~10708, pp.~40--52, SPIE, 2018.

\bibitem{Snowmass2013.Levi}
M.~{Levi}, C.~{Bebek}, T.~{Beers}, R.~{Blum}, R.~{Cahn}, D.~{Eisenstein},
  B.~{Flaugher}, K.~{Honscheid}, R.~{Kron}, O.~{Lahav}, P.~{McDonald},
  N.~{Roe}, D.~{Schlegel}, and {representing the DESI collaboration}, ``{The
  DESI Experiment, a whitepaper for Snowmass 2013},'' {\em arXiv e-prints},
  p.~arXiv:1308.0847, Aug. 2013.

\bibitem{DESI2016a.Science}
{DESI Collaboration}, ``{The DESI Experiment Part I: Science,Targeting, and
  Survey Design},'' {\em arXiv e-prints}, p.~arXiv:1611.00036, Oct. 2016.

\bibitem{DESI2016b.Instr}
{DESI Collaboration}, ``{The DESI Experiment Part II: Instrument Design},''
  {\em arXiv e-prints}, p.~arXiv:1611.00037, Oct. 2016.

\bibitem{DESI2022.KP1.Instr}
{DESI Collaboration}, ``{Overview of the Instrumentation for the Dark Energy
  Spectroscopic Instrument},'' {\em \aj}, vol.~164, p.~207, Nov. 2022.

\bibitem{FocalPlane.Silber.2023}
J.~H. {Silber}, P.~{Fagrelius}, K.~{Fanning}, M.~{Schubnell}, J.~N. {Aguilar},
  S.~{Ahlen}, J.~{Ameel}, O.~{Ballester}, {\em et~al.}, ``{The Robotic
  Multiobject Focal Plane System of the Dark Energy Spectroscopic Instrument
  (DESI)},'' {\em \aj}, vol.~165, p.~9, Jan. 2023.

\bibitem{Corrector.Miller.2023}
T.~N. {Miller}, P.~{Doel}, G.~{Gutierrez}, R.~{Besuner}, {Brooks}, {\em
  et~al.}, ``{The Optical Corrector for the Dark Energy Spectroscopic
  Instrument},'' {\em arXiv e-prints}, p.~arXiv:2306.06310, June 2023.

\bibitem{Spectro.Pipeline.Guy.2023}
J.~{Guy}, S.~{Bailey}, A.~{Kremin}, S.~{Alam}, D.~M. {Alexander}, C.~{Allende
  Prieto}, S.~{BenZvi}, A.~S. {Bolton}, D.~{Brooks}, E.~{Chaussidon}, A.~P.
  {Cooper}, K.~{Dawson}, {de la Macorra}, {\em et~al.}, ``{The Spectroscopic
  Data Processing Pipeline for the Dark Energy Spectroscopic Instrument},''
  {\em \aj}, vol.~165, p.~144, Apr. 2023.

\bibitem{SurveyOps.Schlafly.2023}
E.~F. {Schlafly}, D.~{Kirkby}, D.~J. {Schlegel}, A.~D. {Myers}, A.~{Raichoor},
  K.~{Dawson}, J.~{Aguilar}, C.~{Allende Prieto}, S.~{Bailey}, S.~{BenZvi},
  {\em et~al.}, ``{Survey Operations for the Dark Energy Spectroscopic
  Instrument},'' {\em \aj}, vol.~166, p.~259, Dec. 2023.

\bibitem{LRG.TS.Zhou.2023}
R.~{Zhou}, B.~{Dey}, J.~A. {Newman}, D.~J. {Eisenstein}, K.~{Dawson},
  S.~{Bailey}, A.~{Berti}, J.~{Guy}, T.-W. {Lan}, H.~{Zou}, {\em et~al.},
  ``{Target Selection and Validation of DESI Luminous Red Galaxies},'' {\em
  \aj}, vol.~165, p.~58, Feb. 2023.

\bibitem{DESI2023a.KP1.SV}
{DESI Collaboration}, ``{Validation of the Scientific Program for the Dark
  Energy Spectroscopic Instrument},'' {\em \aj}, vol.~167, p.~62, Feb. 2024.

\bibitem{DESI2023b.KP1.EDR}
{DESI Collaboration}, ``{The Early Data Release of the Dark Energy
  Spectroscopic Instrument},'' {\em arXiv e-prints}, p.~arXiv:2306.06308, June
  2023.

\bibitem{2024arXiv240403000D}
{DESI Collaboration}, ``{DESI 2024 III: Baryon Acoustic Oscillations from
  Galaxies and Quasars},'' {\em arXiv e-prints}, p.~arXiv:2404.03000, Apr.
  2024.

\bibitem{2024arXiv240403001D}
{DESI Collaboration}, ``{DESI 2024 IV: Baryon Acoustic Oscillations from the
  Lyman Alpha Forest},'' {\em arXiv e-prints}, p.~arXiv:2404.03001, Apr. 2024.

\bibitem{2024arXiv240403002D}
{DESI Collaboration}, ``{DESI 2024 VI: Cosmological Constraints from the
  Measurements of Baryon Acoustic Oscillations},'' {\em arXiv e-prints},
  p.~arXiv:2404.03002, Apr. 2024.

\bibitem{laureijs2011euclid}
R.~Laureijs, J.~Amiaux, S.~Arduini, J.-L. Augueres, J.~Brinchmann, R.~Cole,
  M.~Cropper, C.~Dabin, L.~Duvet, A.~Ealet, {\em et~al.}, ``Euclid definition
  study report,'' {\em arXiv preprint arXiv:1110.3193}, 2011.

\bibitem{ivezic2019lsst}
{\v{Z}}.~Ivezi{\'c}, S.~M. Kahn, J.~A. Tyson, B.~Abel, E.~Acosta, R.~Allsman,
  D.~Alonso, Y.~AlSayyad, S.~F. Anderson, J.~Andrew, {\em et~al.}, ``Lsst: from
  science drivers to reference design and anticipated data products,'' {\em The
  Astrophysical Journal}, vol.~873, no.~2, p.~111, 2019.

\bibitem{klein2005blind}
J.~R. Klein and A.~Roodman, ``Blind analysis in nuclear and particle physics,''
  {\em Annu. Rev. Nucl. Part. Sci.}, vol.~55, pp.~141--163, 2005.

\bibitem{conley2006measurement}
A.~Conley, G.~Goldhaber, L.~Wang, G.~Aldering, R.~Amanullah, E.~Commins,
  V.~Fadeyev, G.~Folatelli, G.~Garavini, R.~Gibbons, {\em et~al.},
  ``Measurement of $\omega_m$, $\omega_\lambda$ from a blind analysis of type
  ia supernovae with cmagic: Using color information to verify the acceleration
  of the universe,'' {\em The Astrophysical Journal}, vol.~644, no.~1, p.~1,
  2006.

\bibitem{maccoun2015blind}
R.~MacCoun and S.~Perlmutter, ``Blind analysis: Hide results to seek the
  truth,'' {\em Nature}, vol.~526, no.~7572, pp.~187--189, 2015.

\bibitem{kuijken2015gravitational}
K.~Kuijken, C.~Heymans, H.~Hildebrandt, R.~Nakajima, T.~Erben, J.~T. de~Jong,
  M.~Viola, A.~Choi, H.~Hoekstra, L.~Miller, {\em et~al.}, ``Gravitational
  lensing analysis of the kilo-degree survey,'' {\em Monthly Notices of the
  Royal Astronomical Society}, vol.~454, no.~4, pp.~3500--3532, 2015.

\bibitem{zhang2017blinded}
B.~R. Zhang, M.~J. Childress, T.~M. Davis, N.~V. Karpenka, C.~Lidman, B.~P.
  Schmidt, and M.~Smith, ``A blinded determination of h 0 from low-redshift
  type ia supernovae, calibrated by cepheid variables,'' {\em Monthly Notices
  of the Royal Astronomical Society}, vol.~471, no.~2, pp.~2254--2285, 2017.

\bibitem{abbott2018dark}
T.~M. Abbott, F.~B. Abdalla, A.~Alarcon, J.~Aleksi{\'c}, S.~Allam, S.~Allen,
  A.~Amara, J.~Annis, J.~Asorey, S.~Avila, {\em et~al.}, ``Dark energy survey
  year 1 results: Cosmological constraints from galaxy clustering and weak
  lensing,'' {\em Physical Review D}, vol.~98, no.~4, p.~043526, 2018.

\bibitem{asgari2020kids+}
M.~Asgari, T.~Tr{\"o}ster, C.~Heymans, H.~Hildebrandt, J.~L. van~den Busch,
  A.~H. Wright, A.~Choi, T.~Erben, B.~Joachimi, S.~Joudaki, {\em et~al.},
  ``Kids+ viking-450 and des-y1 combined: Mitigating baryon feedback
  uncertainty with cosebis,'' {\em Astronomy \& Astrophysics}, vol.~634,
  p.~A127, 2020.

\bibitem{muir2020blinding}
J.~Muir, G.~M. Bernstein, D.~Huterer, F.~Elsner, E.~Krause, A.~Roodman,
  S.~Allam, J.~Annis, S.~Avila, K.~Bechtol, {\em et~al.}, ``Blinding multiprobe
  cosmological experiments,'' {\em Monthly Notices of the Royal Astronomical
  Society}, vol.~494, no.~3, pp.~4454--4470, 2020.

\bibitem{sellentin2020blinding}
E.~Sellentin, ``A blinding solution for inference from astronomical data,''
  {\em Monthly Notices of the Royal Astronomical Society}, vol.~492, no.~3,
  pp.~3396--3407, 2020.

\bibitem{fontribera2024inprep}
A.~Font-Ribera {\em et~al.}, ``In prep.,'' 2024.

\bibitem{brieden2020blind}
S.~Brieden, H.~Gil-Mar{\'\i}n, L.~Verde, and J.~L. Bernal, ``Blind observers of
  the sky,'' {\em Journal of Cosmology and Astroparticle Physics}, vol.~2020,
  no.~09, p.~052, 2020.

\bibitem{andrade2024validating}
U.~Andrade, J.~Mena-Fern{\'a}ndez, H.~Awan, A.~Ross, S.~Brieden, J.~Pan,
  A.~de~Mattia, J.~Aguilar, S.~Ahlen, O.~Alves, {\em et~al.}, ``Validating the
  galaxy and quasar catalog-level blinding scheme for the desi 2024 analysis,''
  {\em arXiv preprint arXiv:2404.07282}, 2024.

\bibitem{Alcock:1979mp}
C.~Alcock and B.~Paczynski, ``{An evolution free test for non-zero cosmological
  constant},'' {\em Nature}, vol.~281, pp.~358--359, 1979.

\bibitem{linder2005cosmic}
E.~V. Linder, ``Cosmic growth history and expansion history,'' {\em Physical
  Review D—Particles, Fields, Gravitation, and Cosmology}, vol.~72, no.~4,
  p.~043529, 2005.

\bibitem{linder2007parameterized}
E.~V. Linder and R.~N. Cahn, ``Parameterized beyond-einstein growth,'' {\em
  Astroparticle Physics}, vol.~28, no.~4-5, pp.~481--488, 2007.

\bibitem{eisenstein2007robustness}
D.~J. Eisenstein, H.-J. Seo, and M.~White, ``On the robustness of the acoustic
  scale in the low-redshift clustering of matter,'' {\em The Astrophysical
  Journal}, vol.~664, no.~2, p.~660, 2007.

\bibitem{padmanabhan2009reconstructing}
N.~Padmanabhan, M.~White, and J.~Cohn, ``Reconstructing baryon oscillations: A
  lagrangian theory perspective,'' {\em Physical Review D}, vol.~79, no.~6,
  p.~063523, 2009.

\bibitem{burden2014efficient}
A.~Burden, W.~J. Percival, M.~Manera, A.~J. Cuesta, M.~Vargas~Magana, and
  S.~Ho, ``Efficient reconstruction of linear baryon acoustic oscillations in
  galaxy surveys,'' {\em Monthly Notices of the Royal Astronomical Society},
  vol.~445, no.~3, pp.~3152--3168, 2014.

\bibitem{carter2020impact}
P.~Carter, F.~Beutler, W.~J. Percival, J.~DeRose, R.~H. Wechsler, and C.~Zhao,
  ``The impact of the fiducial cosmology assumption on bao distance scale
  measurements,'' {\em Monthly Notices of the Royal Astronomical Society},
  vol.~494, no.~2, pp.~2076--2089, 2020.

\bibitem{bernal2020robustness}
J.~L. Bernal, T.~L. Smith, K.~K. Boddy, and M.~Kamionkowski, ``Robustness of
  baryon acoustic oscillation constraints for early-universe modifications of
  $\lambda$ cdm cosmology,'' {\em Physical Review D}, vol.~102, no.~12,
  p.~123515, 2020.

\bibitem{sherwin2019impact}
B.~D. Sherwin and M.~White, ``The impact of wrong assumptions in bao
  reconstruction,'' {\em Journal of Cosmology and Astroparticle Physics},
  vol.~2019, no.~02, p.~027, 2019.

\bibitem{Gil-Marin:2014biasgravity}
H.~Gil-Marín, J.~Noreña, L.~Verde, W.~J. Percival, C.~Wagner, M.~Manera, and
  D.~P. Schneider, ``{The power spectrum and bispectrum of SDSS DR11 BOSS
  galaxies – I. Bias and gravity},'' {\em Mon. Not. Roy. Astron. Soc.},
  vol.~451, no.~1, pp.~539--580, 2015.

\bibitem{Gil-Marin:2016sdss}
H.~Gil-Marín, W.~J. Percival, L.~Verde, J.~R. Brownstein, C.-H. Chuang, F.-S.
  Kitaura, S.~A. Rodríguez-Torres, and M.~D. Olmstead, ``{The clustering of
  galaxies in the SDSS-III Baryon Oscillation Spectroscopic Survey: RSD
  measurement from the power spectrum and bispectrum of the DR12 BOSS
  galaxies},'' {\em Mon. Not. Roy. Astron. Soc.}, vol.~465, no.~2,
  pp.~1757--1788, 2017.

\bibitem{pardede2022bispectrum}
K.~Pardede, F.~Rizzo, M.~Biagetti, E.~Castorina, E.~Sefusatti, and P.~Monaco,
  ``Bispectrum-window convolution via hankel transform,'' {\em Journal of
  Cosmology and Astroparticle Physics}, vol.~2022, no.~10, p.~066, 2022.

\bibitem{maksimova2021abacussummit}
N.~A. Maksimova, L.~H. Garrison, D.~J. Eisenstein, B.~Hadzhiyska, S.~Bose, and
  T.~P. Satterthwaite, ``Abacussummit: a massive set of high-accuracy,
  high-resolution n-body simulations,'' {\em Monthly Notices of the Royal
  Astronomical Society}, vol.~508, no.~3, pp.~4017--4037, 2021.

\bibitem{collaboration2018planck}
P.~Collaboration, ``Planck 2018 results. vi. cosmological parameters,'' 2018.

\bibitem{tegmark1997measuring}
M.~Tegmark, ``Measuring cosmological parameters with galaxy surveys,'' {\em
  Physical Review Letters}, vol.~79, no.~20, p.~3806, 1997.

\bibitem{variuinprep}
A.~Variu {\em et~al.}, ``In prep.,'' 2024.

\bibitem{white2014mock}
M.~White, J.~L. Tinker, and C.~K. McBride, ``Mock galaxy catalogues using the
  quick particle mesh method,'' {\em Monthly Notices of the Royal Astronomical
  Society}, vol.~437, no.~3, pp.~2594--2606, 2014.

\bibitem{chuang2015ezmocks}
C.-H. Chuang, F.-S. Kitaura, F.~Prada, C.~Zhao, and G.~Yepes, ``Ezmocks:
  extending the zel'dovich approximation to generate mock galaxy catalogues
  with accurate clustering statistics,'' {\em Monthly Notices of the Royal
  Astronomical Society}, vol.~446, no.~3, pp.~2621--2628, 2015.

\bibitem{HGMeboss}
H.~{Gil-Mar{\'\i}n}, J.~E. {Bautista}, R.~{Paviot}, M.~{Vargas-Maga{\~n}a},
  S.~{de la Torre}, S.~{Fromenteau}, S.~{Alam}, S.~{{\'A}vila}, E.~{Burtin},
  C.-H. {Chuang}, K.~S. {Dawson}, J.~{Hou}, A.~{de Mattia}, F.~G. {Mohammad},
  E.-M. {M{\"u}ller}, S.~{Nadathur}, R.~{Neveux}, W.~J. {Percival},
  A.~{Raichoor}, M.~{Rezaie}, A.~J. {Ross}, G.~{Rossi}, V.~{Ruhlmann-Kleider},
  A.~{Smith}, A.~{Tamone}, J.~L. {Tinker}, R.~{Tojeiro}, Y.~{Wang}, G.-B.
  {Zhao}, C.~{Zhao}, J.~{Brinkmann}, J.~R. {Brownstein}, P.~D. {Choi},
  S.~{Escoffier}, A.~{de la Macorra}, J.~{Moon}, J.~A. {Newman}, D.~P.
  {Schneider}, H.-J. {Seo}, and M.~{Vivek}, ``{The Completed SDSS-IV extended
  Baryon Oscillation Spectroscopic Survey: measurement of the BAO and growth
  rate of structure of the luminous red galaxy sample from the anisotropic
  power spectrum between redshifts 0.6 and 1.0},'' {\em \mnras}, vol.~498,
  pp.~2492--2531, Oct. 2020.

\bibitem{novell2023geofpt}
S.~Novell-Masot, D.~Gualdi, H.~Gil-Mar{\'\i}n, and L.~Verde, ``Geo-fpt: a model
  of the galaxy bispectrum at mildly non-linear scales,'' {\em Journal of
  Cosmology and Astroparticle Physics}, vol.~2023, no.~11, p.~044, 2023.

\bibitem{novell2023approximations}
S.~Novell-Masot, H.~Gil-Mar{\'\i}n, and L.~Verde, ``On approximations of the
  redshift-space bispectrum and power spectrum multipoles covariance matrix,''
  {\em arXiv preprint arXiv:2306.03137}, 2023.

\bibitem{Crocce_2006}
M.~Crocce and R.~Scoccimarro, ``Renormalized cosmological perturbation
  theory,'' {\em Physical Review D}, vol.~73, mar 2006.

\bibitem{gil-marin_power_2015}
H.~Gil-Marín, J.~Noreña, L.~Verde, W.~J. Percival, C.~Wagner, M.~Manera, and
  D.~P. Schneider, ``The power spectrum and bispectrum of {SDSS} {DR11} {BOSS}
  galaxies – {I}. {Bias} and gravity,'' {\em Monthly Notices of the Royal
  Astronomical Society}, vol.~451, pp.~539--580, July 2015.

\bibitem{Scoccimarro:1999ed}
R.~Scoccimarro, H.~M.~P. Couchman, and J.~A. Frieman, ``{The Bispectrum as a
  Signature of Gravitational Instability in Redshift-Space},'' {\em Astrophys.
  J.}, vol.~517, pp.~531--540, 1999.

\bibitem{baldauf2012evidence}
T.~Baldauf, U.~Seljak, V.~Desjacques, and P.~McDonald, ``Evidence for quadratic
  tidal tensor bias from the halo bispectrum,'' {\em Physical Review D},
  vol.~86, no.~8, p.~083540, 2012.

\bibitem{saito2014understanding}
S.~Saito, T.~Baldauf, Z.~Vlah, U.~Seljak, T.~Okumura, and P.~McDonald,
  ``Understanding higher-order nonlocal halo bias at large scales by combining
  the power spectrum with the bispectrum,'' {\em Physical Review D}, vol.~90,
  no.~12, p.~123522, 2014.

\bibitem{Brieden_ptchallenge}
S.~{Brieden}, H.~{Gil-Mar{\'\i}n}, and L.~{Verde}, ``{PT challenge: validation
  of ShapeFit on large-volume, high-resolution mocks},'' {\em \jcap},
  vol.~2022, p.~005, June 2022.

\bibitem{Hartlap:2006kj}
J.~Hartlap, P.~Simon, and P.~Schneider, ``{Why your model parameter confidences
  might be too optimistic: Unbiased estimation of the inverse covariance
  matrix},'' {\em Astron. Astrophys.}, vol.~464, p.~399, 2007.

\bibitem{Sellentin:2015waz}
E.~Sellentin and A.~F. Heavens, ``{Parameter inference with estimated
  covariance matrices},'' {\em Mon. Not. Roy. Astron. Soc.}, vol.~456, no.~1,
  pp.~L132--L136, 2016.

\bibitem{Gualdi:2020aniso}
D.~Gualdi and L.~Verde, ``{Galaxy redshift-space bispectrum: the Importance of
  Being Anisotropic},'' {\em JCAP}, vol.~06, p.~041, 2020.

\bibitem{eisenstein2005detection}
D.~J. Eisenstein, I.~Zehavi, D.~W. Hogg, R.~Scoccimarro, M.~R. Blanton, R.~C.
  Nichol, R.~Scranton, H.-J. Seo, M.~Tegmark, Z.~Zheng, {\em et~al.},
  ``Detection of the baryon acoustic peak in the large-scale correlation
  function of sdss luminous red galaxies,'' {\em The Astrophysical Journal},
  vol.~633, no.~2, p.~560, 2005.

\bibitem{bautista2021completed}
J.~E. Bautista, R.~Paviot, M.~Vargas~Magana, S.~de~La~Torre, S.~Fromenteau,
  H.~Gil-Mar{\'\i}n, A.~J. Ross, E.~Burtin, K.~S. Dawson, J.~Hou, {\em et~al.},
  ``The completed sdss-iv extended baryon oscillation spectroscopic survey:
  measurement of the bao and growth rate of structure of the luminous red
  galaxy sample from the anisotropic correlation function between redshifts 0.6
  and 1,'' {\em Monthly Notices of the Royal Astronomical Society}, vol.~500,
  no.~1, pp.~736--762, 2021.

\bibitem{carrasco2012effective}
J.~J.~M. Carrasco, M.~P. Hertzberg, and L.~Senatore, ``The effective field
  theory of cosmological large scale structures,'' {\em Journal of High Energy
  Physics}, vol.~2012, no.~9, pp.~1--40, 2012.

\bibitem{nishimichi2020blinded}
T.~Nishimichi, G.~D’Amico, M.~M. Ivanov, L.~Senatore, M.~Simonovi{\'c},
  M.~Takada, M.~Zaldarriaga, and P.~Zhang, ``Blinded challenge for precision
  cosmology with large-scale structure: results from effective field theory for
  the redshift-space galaxy power spectrum,'' {\em Physical Review D},
  vol.~102, no.~12, p.~123541, 2020.

\bibitem{chudaykin2020nonlinear}
A.~Chudaykin, M.~M. Ivanov, O.~H. Philcox, and M.~Simonovi{\'c}, ``Nonlinear
  perturbation theory extension of the boltzmann code class,'' {\em Physical
  Review D}, vol.~102, no.~6, p.~063533, 2020.

\bibitem{colas2020efficient}
T.~Colas, G.~d'Amico, L.~Senatore, P.~Zhang, and F.~Beutler, ``Efficient
  cosmological analysis of the sdss/boss data from the effective field theory
  of large-scale structure,'' {\em Journal of Cosmology and Astroparticle
  Physics}, vol.~2020, no.~06, p.~001, 2020.

\bibitem{ivanov2020cosmological}
M.~M. Ivanov, M.~Simonovi{\'c}, and M.~Zaldarriaga, ``Cosmological parameters
  from the boss galaxy power spectrum,'' {\em Journal of Cosmology and
  Astroparticle Physics}, vol.~2020, no.~05, p.~042, 2020.

\bibitem{d2022boss}
G.~D'Amico, Y.~Donath, M.~Lewandowski, L.~Senatore, and P.~Zhang, ``The boss
  bispectrum analysis at one loop from the effective field theory of
  large-scale structure,'' {\em arXiv preprint arXiv:2206.08327}, 2022.

\bibitem{Foreman_Mackey_2013}
D.~Foreman-Mackey, D.~W. Hogg, D.~Lang, and J.~Goodman, ``emcee: The {MCMC}
  hammer,'' {\em Publications of the Astronomical Society of the Pacific},
  vol.~125, pp.~306--312, mar 2013.

\bibitem{gsl}
M.~Galassi, J.~Davies, J.~Theiler, B.~Gough, G.~Jungman, P.~Alken, M.~Booth,
  F.~Rossi, and R.~Ulerich, {\em GNU scientific library}.
\newblock Network Theory Limited Godalming, 2002.

\bibitem{scipy}
{SciPy 1.0 Contributors}, ``{{SciPy} 1.0: Fundamental Algorithms for Scientific
  Computing in Python},'' {\em Nature Methods}, vol.~17, pp.~261--272, 2020.

\bibitem{numpy}
C.~R. Harris {\em et~al.}, ``Array programming with {NumPy},'' {\em Nature},
  vol.~585, pp.~357--362, Sept. 2020.

\bibitem{getdist}
A.~Lewis, ``Getdist: a python package for analysing monte carlo samples,'' {\em
  arXiv preprint arXiv:1910.13970}, 2019.

\bibitem{astropy}
{Astropy Collaboration}, ``{The Astropy Project: Sustaining and Growing a
  Community-oriented Open-source Project and the Latest Major Release (v5.0) of
  the Core Package},'' {\em \apj}, vol.~935, p.~167, Aug. 2022.

\bibitem{matplotlib}
J.~D. Hunter, ``Matplotlib: A 2d graphics environment,'' {\em Computing in
  Science \& Engineering}, vol.~9, no.~3, pp.~90--95, 2007.

\bibitem{Taruya_2010}
A.~Taruya, T.~Nishimichi, and S.~Saito, ``Baryon acoustic oscillations in 2d:
  Modeling redshift-space power spectrum from perturbation theory,'' {\em
  Physical Review D}, vol.~82, sep 2010.

\bibitem{Nishimichi_2011}
T.~Nishimichi and A.~Taruya, ``Baryon acoustic oscillations in 2d. {II}.
  redshift-space halo clustering in n-body simulations,'' {\em Physical Review
  D}, vol.~84, aug 2011.

\bibitem{beutler2014clustering}
F.~Beutler, S.~Saito, H.-J. Seo, J.~Brinkmann, K.~S. Dawson, D.~J. Eisenstein,
  A.~Font-Ribera, S.~Ho, C.~K. McBride, F.~Montesano, {\em et~al.}, ``The
  clustering of galaxies in the sdss-iii baryon oscillation spectroscopic
  survey: testing gravity with redshift space distortions using the power
  spectrum multipoles,'' {\em Monthly Notices of the Royal Astronomical
  Society}, vol.~443, no.~2, pp.~1065--1089, 2014.

\bibitem{Jackson_1972}
J.~C. Jackson, ``A critique of rees{\textquotesingle}s theory of primordial
  gravitational radiation,'' {\em Monthly Notices of the Royal Astronomical
  Society}, vol.~156, pp.~1P--5P, feb 1972.

\bibitem{Verde1998}
L.~Verde, A.~F. Heavens, S.~Matarrese, and L.~Moscardini, ``{Large-scale bias
  in the Universe - II. Redshift-space bispectrum},'' {\em Mon. Not. R. Astron.
  Soc.}, vol.~300, no.~3, pp.~747--756, 1998.

\bibitem{Gil-Marin:2014theory}
H.~Gil-Marín, C.~Wagner, J.~Noreña, L.~Verde, and W.~Percival, ``{Dark matter
  and halo bispectrum in redshift space: theory and applications},'' {\em
  JCAP}, vol.~12, p.~029, 2014.

\bibitem{gil-marin_clustering_2017}
H.~Gil-Marín, W.~J. Percival, L.~Verde, J.~R. Brownstein, C.-H. Chuang, F.-S.
  Kitaura, S.~A. Rodríguez-Torres, and M.~D. Olmstead, ``The clustering of
  galaxies in the {SDSS}-{III} {Baryon} {Oscillation} {Spectroscopic} {Survey}:
  {RSD} measurement from the power spectrum and bispectrum of the {DR12} {BOSS}
  galaxies,'' {\em Monthly Notices of the Royal Astronomical Society},
  vol.~465, pp.~1757--1788, Feb. 2017.

\bibitem{hamilton1992measuring}
A.~Hamilton, ``Measuring omega and the real correlation function from the
  redshift correlation function,'' {\em The Astrophysical Journal}, vol.~385,
  pp.~L5--L8, 1992.

\bibitem{cole1994fourier}
S.~Cole, K.~B. Fisher, and D.~H. Weinberg, ``Fourier analysis of redshift-space
  distortions and the determination of $\omega$,'' {\em Monthly Notices of the
  Royal Astronomical Society}, vol.~267, no.~3, pp.~785--799, 1994.

\end{thebibliography}

\end{document}